\documentclass[aps,prd,10pt,twocolumn,superscriptaddress,amssymb,amsfonts,amsmath,nofootinbib,preprintnumbers]{revtex4-2}
\usepackage{graphicx,slashed,xcolor,multirow,bbold,mathtools,sidecap,tikz,bm,ulem,enumitem,booktabs,array}
\usepackage[colorlinks=true,linktoc=page,linkcolor=purple,citecolor=teal,urlcolor=magenta]{hyperref}
\usepackage[compat=1.1.0]{tikz-feynman}
\usepackage{siunitx,adjustbox,comment}
\setenumerate{label=$\roman*)$,itemsep=2pt,parsep=2pt,topsep=2pt,partopsep=2pt,leftmargin=18pt}
\definecolor{lime}{HTML}{A6CE39}
\DeclareRobustCommand{\orcidicon}{\hspace{-2.1mm}
\begin{tikzpicture}
\draw[lime,fill=lime] (0,0.0) circle [radius=0.13] node[white] {{\fontfamily{qag}\selectfont \tiny ID}}; \draw[white,fill=white] (-0.0525,0.095) circle [radius=0.007]; 
\end{tikzpicture} \hspace{-3.5mm} }
\foreach \x in {A, ..., Z} {\expandafter\xdef\csname orcid\x\endcsname{\noexpand\href{https://orcid.org/\csname orcidauthor\x\endcsname} {\noexpand\orcidicon}}}

\let\emph\textit

\begin{document}

\preprint{TTP25-055, P3H-25-108}

\title{Precision Higgs Boson Probe of Type-II Seesaw Models}

\author{Saiyad Ashanujjaman\orcidA}
\email{saiyad.ashanujjaman@kit.edu}
\affiliation{Institut f\"ur Theoretische Teilchenphysik, Karlsruhe Institute of Technology, Engesserstra\ss e 7, D-76128 Karlsruhe, Germany}
\affiliation{Institut f\"ur Astroteilchenphysik, Karlsruhe Institute of Technology, Hermann-von-Helmholtz-Platz 1, D-76344 Eggenstein-Leopoldshafen, Germany}

\author{P.~S.~Bhupal Dev\orcidB}
\email{bdev@wustl.edu}
\affiliation{Department of Physics and McDonnell Center for the Space Sciences, Washington University, Saint Louis, MO 63130, USA}

\author{Jihong Huang\orcidC}
\email{huangjh@ihep.ac.cn}
\affiliation{Institute of High Energy Physics, Chinese Academy of Sciences, Beijing 100049, China}
\affiliation{School of Physical Sciences, University of Chinese Academy of Sciences, Beijing 100049, China}

\author{Shun Zhou\orcidD}
\email{zhoush@ihep.ac.cn}
\affiliation{Institute of High Energy Physics, Chinese Academy of Sciences, Beijing 100049, China}
\affiliation{School of Physical Sciences, University of Chinese Academy of Sciences, Beijing 100049, China}
                                                                             
\begin{abstract}
Despite direct searches at the LHC excluding tripletlike Higgs bosons up to several hundred GeV over much of the type-II seesaw model parameter space, parts of it---most notably those featuring ``cascade decays'' of the charged Higgs bosons into their neutral partners and off-shell $W$ bosons---still remain unconstrained. Meanwhile, measurements of the diphoton signal strength of the Standard Model (SM) Higgs boson---potentially modified by loop contributions from tripletlike Higgs states---are in good agreement with the SM expectation, with combined experimental uncertainties currently at approximately 8\%. Given the trend in previous measurements, it is expected that future precision Higgs measurements at the HL-LHC and a future lepton collider such as the Circular Electron Positron Collider, Future Circular Collider, or Muon Collider will be consistent the standard diphoton signal strength, albeit with significantly reduced uncertainties, down to about 0.7\%. Presuming this and considering all relevant constraints, we explore whether such increasingly precise diphoton measurements can indirectly probe the parameter space that currently evades direct searches. We find that subpercent-level determinations of the diphoton rate will decisively probe a substantial fraction of this otherwise elusive region.
\end{abstract}

\maketitle

\section{Introduction}
\label{sec:intro}
The observation of nonzero neutrino masses and mixing  provides a strong impetus for going beyond the Standard Model (SM). A particularly elegant resolution is provided by the Weinberg operator-induced seesaw mechanism~\cite{Weinberg:1979sa}. Among its three canonical ultraviolet completions~\cite{Ma:1998dn}, the type-II seesaw model~\cite{Konetschny:1977bn,Schechter:1980gr,Cheng:1980qt,Magg:1980ut,Lazarides:1980nt,Mohapatra:1980yp}, which augments the SM with an ${\rm SU(2)}^{}_{\rm L}$-triplet scalar field carrying hypercharge $Y=2$, has attracted considerable interest due to its distinctive phenomenological signatures and potential testability at current and future colliders~\cite{Huitu:1996su, Gunion:1996pq, Chakrabarti:1998qy, Datta:1999nc, Muhlleitner:2003me, Chun:2003ej, Akeroyd:2005gt,  Han:2007bk, 
Garayoa:2007fw, 
Kadastik:2007yd, 
Akeroyd:2007zv,
FileviezPerez:2008jbu, 
delAguila:2008cj, Akeroyd:2009hb, 
Rodejohann:2010jh, Akeroyd:2010ip, Rodejohann:2010bv, Melfo:2011nx, Aoki:2011pz, Akeroyd:2011zza, Arhrib:2011uy, 
Akeroyd:2012nd,
Chiang:2012dk,  Chun:2012zu, 
Chun:2012jw, 
Aoki:2012jj,  Arbabifar:2012bd,
Dev:2013ff,
Kanemura:2013vxa, 
Chun:2013vma,
delAguila:2013mia,  
kang:2014jia, Kanemura:2014goa, Kanemura:2014ipa, Han:2015hba, 
Han:2015sca,  
Das:2016bir, 
Blunier:2016peh, 
Mitra:2016wpr, 
Babu:2016rcr,
Nomura:2017abh, 
Dev:2017ouk, Ghosh:2017pxl, 
Agrawal:2018pci, 
Dev:2018vpr, 
Dev:2018sel, 
Crivellin:2018ahj, 
Dev:2018tox, Antusch:2018svb, 
Dev:2019hev, 
Primulando:2019evb, 
deMelo:2019asm, 
Rahili:2019ixf,
Padhan:2019jlc,
Chun:2019hce, 
Fuks:2019clu,
Gluza:2020qrt, Bandyopadhyay:2020mnp, 
Yang:2021skb, Ashanujjaman:2021txz, 
Heeck:2022fvl,
Ashanujjaman:2022tdn,
Bahl:2022gqg,
Cheng:2022hbo,
Butterworth:2022dkt,
Ashanujjaman:2022ofg,
Li:2023ksw, 
Xu:2023ene, 
Maharathy:2023dtp,
Ashanujjaman:2023tlj,  Fridell:2023gjx,
Lichtenstein:2023iut,
Dev:2023nha,
Jia:2024wqi, Ducu:2024xxf} (for reviews, see Refs.~\cite{Deppisch:2015qwa, Cai:2017mow}), as well as at low-energy experiments~\cite{Kakizaki:2003jk, Akeroyd:2009nu,Dinh:2012bp, Dev:2017ouk, 
Dev:2018sel, Crivellin:2018ahj,
Barrie:2022ake}. This model introduces several new Higgs bosons, namely, the doubly charged ($H^{\pm\pm}$), singly charged ($H^\pm$), and $CP$-even  and $CP$-odd neutral ($H^0$ and $A^0$) Higgs bosons. Direct searches at the LEP and LHC experiments have put stringent constraints on these new scalars. LEP experiments (OPAL and DELPHI) have set a lower-mass limit of about 100 GeV for $H^{\pm\pm}$ decaying into $\ell^\pm\ell^\pm$ ($\ell=e,\mu$)~\cite{OPAL:2001luy, DELPHI:2002bkf}. For $H^\pm$ decaying into $cs$ or $\tau\nu$, a lower limit of about 80 GeV has been set by ALEPH, DELPHI, L3, and OPAL ~\cite{LEPHiggsWorkingGroupforHiggsbosonsearches:2001ogs}. The OPAL experiment also placed a limit of 55 GeV on the neutral scalar states $H^0$ and $A^0$ decaying into $\nu\bar{\nu}$~\cite{Datta:1999nc}. At the LHC, the ATLAS Collaboration has set a lower limit of 1020 GeV for $H^{\pm\pm}$ decaying into $\ell^\pm\ell^\pm$~\cite{ATLAS:2014kca, ATLAS:2017xqs, ATLAS:2022pbd}, while CMS has placed a limit of 535 GeV for decays exclusively into $\tau^\pm \tau^\pm$ ~\cite{CMS:2017pet}. Also, ATLAS has excluded $H^{\pm\pm}$ decaying into $W^\pm W^\pm$ in the 200–350 GeV mass range \cite{ATLAS:2021jol}.\footnote{Ref.~\cite{Ashanujjaman:2021txz} has estimated an improved exclusion range of 200–400 GeV by reinterpreting some of these LHC searches.}

Broadly speaking, the phenomenology, and hence the exclusion limits, of the type-II seesaw model depends on three parameters: the doubly charged Higgs mass $m_{H^{\pm \pm}}$, the mass splitting $\Delta m=m_{H^{\pm \pm}}-m_{H^\pm}$, and the triplet vacuum expectation value (VEV) $v_\Delta$~\cite{FileviezPerez:2008jbu, Aoki:2011pz, Ashanujjaman:2021txz}. The regions with $\Delta m \lesssim {\cal O}(1)$ GeV  are dominated by the ``golden decays'' of $H^{\pm \pm}$ to $\ell^\pm \ell^\pm$ (or $W^\pm W^\pm$) for $v_\Delta$ smaller (or larger) than ${\cal O}(0.1)~{\rm MeV}$, and therefore are highly constrained by the experimental searches mentioned above. In contrast, the region with larger $\Delta m \gtrsim \mathcal{O}(1)$ GeV is dominated by the ``cascade decays'' of $H^{\pm\pm}$ and $H^\pm$ to $H^0/A^0$ and off-shell $W$-bosons, with $H^0/A^0$ subsequently decaying into $\nu\bar\nu$ or $b\bar{b},t\bar{t},ZZ,Zh,hh$ depending on $v_\Delta$. The presence of off-shell $W$ bosons and multiple decay steps leads to soft visible objects, missing energy, and large SM backgrounds, rendering conventional LHC searches ineffective. Consequently, this region remains largely unconstrained by direct searches to date~\cite{Ashanujjaman:2021txz, Ashanujjaman:2022tdn, Ashanujjaman:2023tlj}. Moreover, low-mass $H^{\pm\pm}$ (84--200 GeV) decaying into $W^\pm W^\pm$ are yet to be excluded~\cite{Kanemura:2014goa, Ashanujjaman:2022ofg}. In this Letter, we show for the first time, to the best of our knowledge, that indirect probes---such as precision measurements of the SM Higgs boson properties at the LHC and future lepton colliders like Circular Electron Positron Collider (CEPC), Future Circular Collider (FCC-ee) or MuC---could provide a powerful complementary handle on this otherwise elusive region of parameter space. 

In particular, the new charged Higgs bosons $H^\pm$ and $H^{\pm\pm}$, if relatively light, can significantly contribute to the loop-induced SM Higgs decay into diphotons, $h\to\gamma\gamma$~\cite{Chun:2012jw, Dev:2013ff, Das:2016bir, Lichtenstein:2023vza}. This, in turn, imposes nontrivial constraints on regions of the model parameter space that would otherwise evade direct searches. Notably, the LHC measurements of the corresponding signal strength $\mu_{h\to\gamma\gamma}$ are in good agreement with the SM prediction~\cite{ATLAS:2016neq, CMS:2022dwd, ATLAS:2022vkf}, with the current experimental uncertainty of approximately 10\%~\cite{CMS:2021kom, ATLAS:2022tnm}, down to about 8\% when combined with the Tevatron data~\cite{Heo:2024cif}.  Even though recent determinations lie somewhat above the SM expectation, the pattern of past measurements suggests that upcoming results from the HL-LHC~\cite{Cepeda:2019klc}, and a future lepton collider  such as  CEPC~\cite{CEPCStudyGroup:2018ghi}, FCC-ee~\cite{FCC:2018byv}, or MuC~\cite{Accettura:2023ked}, will likely consolidate the SM prediction for the SM Higgs-to-diphoton signal strength, albeit with significantly improved precision---potentially reducing uncertainties to as low as 1.3\% (0.7\%) when combining HL-LHC measurements with those from CEPC or FCC-ee (MuC)~\cite{deBlas:2019rxi, Castelli:2025mqk}. Presuming such a scenario, and taking into account all relevant constraints, we explore in this Letter how future precision measurements of the Higgs diphoton signal could provide a complementary probe of the hitherto unconstrained elusive regions of the type-II seesaw model parameter space.  


\section{The model and Current Constraints}
\label{sec:model}
In the type-II seesaw model~\cite{Konetschny:1977bn,Cheng:1980qt,Schechter:1980gr,Magg:1980ut,Lazarides:1980nt,Mohapatra:1980yp}, the SM is augmented with an ${\rm SU}(2)^{}_{\rm L}$-triplet scalar field with $Y=2$:
\begin{align}
\Delta = \begin{pmatrix} \Delta^+/\sqrt{2} & \Delta^{++} \\ \Delta^0 & -\Delta^+/\sqrt{2} \end{pmatrix} \;.
\end{align}
The most general scalar potential is given by
\begin{align}
V =& -m_\Phi^2{\Phi^\dagger \Phi} + \frac{\lambda}{4}(\Phi^\dagger \Phi)^2 + m_\Delta^2{\rm Tr}(\Delta^{\dagger}{\Delta}) \, \nonumber
\\
& +[\mu(\Phi^T{i}\sigma^2\Delta^\dagger \Phi)+{\rm H.c.}] + \lambda_1(\Phi^\dagger \Phi){\rm Tr}(\Delta^{\dagger}{\Delta}) \, 
\\
& +\lambda_2[{\rm Tr}(\Delta^{\dagger}{\Delta})]^2 + \lambda_3{\rm Tr}[(\Delta^{\dagger}{\Delta})^2] + \lambda_4{\Phi^\dagger \Delta \Delta^\dagger \Phi}\;, \nonumber 
\end{align}
where $\Phi$ is the SM Higgs doublet; $m_\Phi^2, m_\Delta^2$ and $\mu$ are the mass parameters, and $\lambda$ and $\lambda_i$ ($i=1,\dots,4$) are the dimensionless quartic couplings. The neutral components of $\Phi$ and $\Delta$ procure respective VEVs $v_\Phi/\sqrt{2}$ and $v_\Delta/\sqrt{2}$, satisfying $\sqrt{v_\Phi^2+2v_\Delta^2} = v \approx 246$~GeV. For a detailed description of the scalar potential and its dynamical features, see Ref.~\cite{Arhrib:2011uy}. The electroweak symmetry breaking gives rise to several physical states: two $CP$-even states ($h$ and $H^0$), two $CP$-odd states ($G^0$ and $A^0$), two singly charged states ($G^\pm$ and $H^\pm$), and a doubly charged state $H^{\pm \pm}$. Here, $G^0$ and $G^\pm$ are the {\it would-be} Nambu-Goldstone bosons, and $h$ is identified as the 125 GeV Higgs boson observed at the LHC. All the scalar potential parameters can be traded in terms of the physical masses $m_h, m_{H^0}, m_{A^0}, m_{H^\pm}, m_{H^{\pm \pm}}$, the VEVs $v_\Phi,v_\Delta$, and the rotation angle $\alpha$ of the $CP$-even Higgs states~\cite{Arhrib:2011uy}. For $v_\Delta^2 \ll v_\Phi^2$ (as required by electroweak precision constraints, see below), the tripletlike Higgs states follow the sum rule
\begin{align}
m_{H^{\pm\pm}}^2-m_{H^\pm}^2 \approx m_{H^\pm}^2 - m_{H^0/A^0}^2 \approx -\frac{\lambda_4}{4}v_\Phi^2 \; .
\label{eq:sum}
\end{align}

The model also features a Yukawa interaction term  with the SM lepton doublet: 
\begin{align}
    -{\cal L}_{\rm int}\supset Y_{\alpha\beta} L^T_\alpha C i \sigma^2 \Delta L_\beta +{\rm H.c.}\; ,
    \label{eq:Yuk}
\end{align}
where $C$ is the charge conjugation matrix, $\sigma^2$ is the second Pauli matrix, and $(\alpha,\beta)$ denote the lepton flavor index. This Yukawa term, together with the $\mu$-term in the scalar potential, breaks lepton number when $\Delta$ acquires a VEV, thereby generating tree-level Majorana neutrino masses:
\begin{align}
m_\nu=\sqrt{2}Y v_\Delta \; .
\end{align}

\begin{figure}[t!]
\centering
\includegraphics[width=0.4\textwidth]{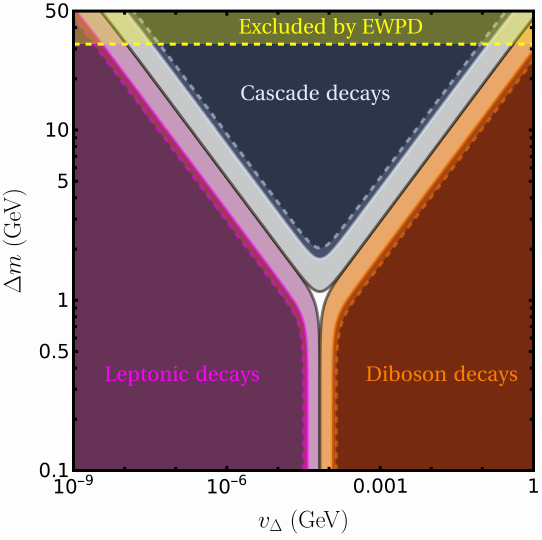}
\caption{Decay phase diagram of the doubly charged Higgs bosons in the $v_\Delta$--$\Delta m$ parameter space. Dashed (solid) contours in magenta, orange, and light blue represent 95\% (90\%) branching ratios, while black solid contours indicate 50\% branching ratios for the corresponding decay modes. The region shown in blue remains unconstrained by the direct searches thus far. For definiteness, we set $m_{H^{\pm\pm}} = 400$~GeV. The yellow-shaded region above the dashed horizontal line is excluded by the current EWPD at 95\% CL. For the other tripletlike Higgs states, the decay phase diagrams are qualitatively similar. For the $\Delta m < 0$ scenario, the region with $\Delta m \lesssim -40$\,GeV is excluded by EWPD.}
\label{fig:decay}
\end{figure}

The tripletlike Higgs bosons can decay either into a pair of SM particles or into a lighter triplet partner accompanied by an off-shell $W/Z$ boson. In broad terms, these decays can be categorized into three classes: $(i)$ leptonic decays, $(ii)$ diboson decays, and $(iii)$ cascade decays, in which the parent Higgs decays into a lighter partner and an off-shell $W/Z$ boson. The branching ratios for these decay modes are highly sensitive to two key parameters: the mass splitting $\Delta m = m_{H^{\pm\pm}} - m_{H^\pm}$ and the triplet VEV $v_\Delta$ (see, for instance, Ref.~\cite{Ashanujjaman:2021txz}). In Fig.~\ref{fig:decay}, we present the decay phase diagram of tripletlike Higgs bosons (see also Ref.~\cite{Melfo:2011nx}), illustrating regions where different decay categories dominate in the $v_\Delta$--$\Delta m$ parameter space. For definiteness, we show only the case of the doubly charged Higgs boson $H^{\pm\pm}$ with a mass of 400~GeV. Similar decay patterns are observed for the other tripletlike Higgs states, though these are not shown for brevity. 

As discussed in Sec.~\ref{sec:intro}, the parameter space dominated by cascade decays of $H^{\pm\pm}$ and $H^\pm$ to $H^0/A^0$ plus off-shell $W$ bosons---depicted in blue in Fig.~\ref{fig:decay}---remains largely unconstrained by the direct searches thus far; see Fig.~\ref{fig:limit} for a summary of the current exclusion limit on $m_{H^{\pm\pm}}$ in the $v_\Delta-\Delta m$ plane. In what follows, we briefly outline the relevant constraints on the model:

\begin{figure}[t!]
\centering
\includegraphics[width=0.49\textwidth]{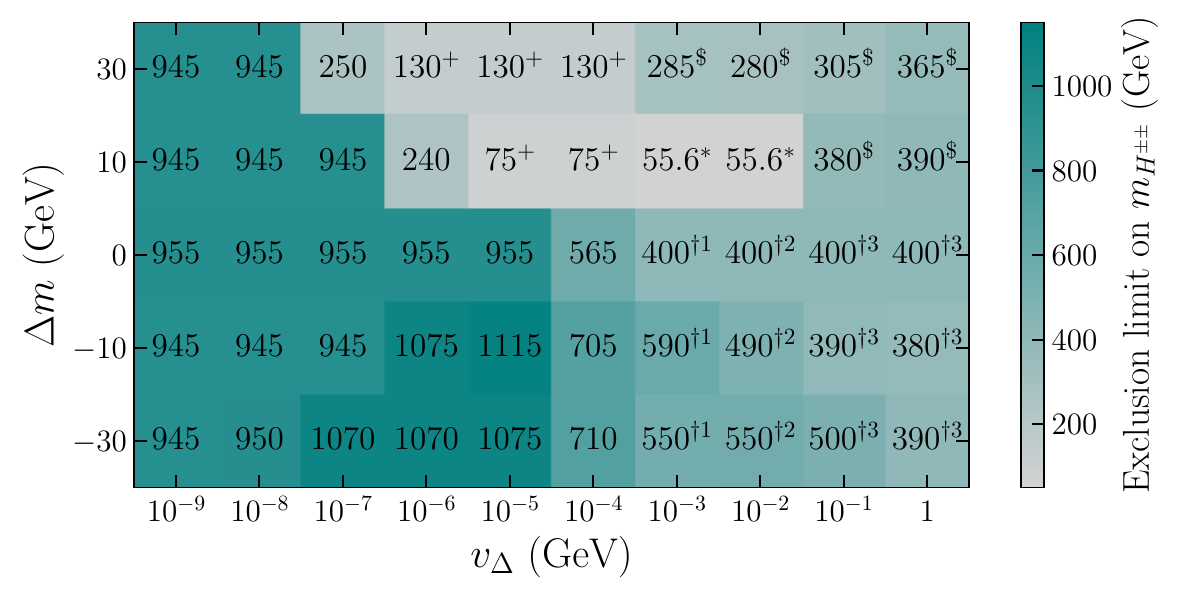}
\caption{Estimated 95\% CL lower limits on $m_{H^{\pm\pm}}$ in the $v_\Delta$--$\Delta m$ parameter space. Entries marked with superscript `$\dagger 1$',  (`$\dagger 2$') [`$\dagger 3$'] and `\$'  respectively denote region where masses of 150--200 (95--200) [80--200]\,GeV and 55.6--200\,GeV are not excluded yet; those with a superscript `+' (`*') refer to limits derived from LEP monophoton searches ($Z$-pole width bound on $Z\to H^+H^-$).}
\label{fig:limit}
\end{figure}

\begin{enumerate}
\item {\it Theoretical constraints:} Tree-level perturbative unitarity and boundedness-from-below conditions on the scalar couplings restrict the theoretically viable parameter space within the perturbative regime. These constraints have been extensively studied; see, e.g., Refs.~\cite{Arhrib:2011uy, Arbabifar:2012bd, Chun:2012jw}. 
\item {\it Lepton flavor violation:} The Yukawa interaction \eqref{eq:Yuk} leads to flavor-changing lepton decays such as $\mu \to e\gamma$, $\mu \to eee$, and muonium-antimuonium oscillation. This also results in enhanced contributions to $e^-e^+ \to \ell^-\ell^+$ cross sections. Upper limits on these processes, combined with neutrino oscillation data, impose a {\it lower} bound on $v_\Delta \gtrsim \mathcal{O}({\rm eV})$ for $m_{H^{\pm\pm}} \sim {\cal O}({\rm TeV})$~\cite{Kakizaki:2003jk, Akeroyd:2009nu, Dinh:2012bp, Dev:2018sel, Dev:2018vpr}. The exact limit depends on the lightest neutrino mass and the mass hierarchy.
\item {\it Collider Limits:} In Fig.~\ref{fig:limit}, we summarize the limits from collider searches in the $v_\Delta$--$\Delta m$ parameter space. The entries indicate the 95\% CL lower limits on  $m_{H^{\pm\pm}}$, as obtained in Refs.~\cite{Datta:1999nc, ATLAS:2014kca, Ashanujjaman:2021txz}. Entries marked with superscripts `$\dagger 1$'  (`$\dagger 2$')   [`$\dagger 3$'] and `\$'  respectively  denote regions where masses of 150--200 (95--200)  [80--200]\,GeV and 55.6--200\,GeV are not yet excluded. Those with a superscript `+' (`*') refer to limits derived from monophoton searches ($Z$-pole width bound on $Z\to H^+H^-$) at LEP~\cite{Datta:1999nc}. In regions where $H^\pm$ decays to $\ell^\pm \nu$ or $\tau\nu$, limits from slepton searches at LEP and the LHC, lepton universality in $W$ decays from LEP, as well as from $\tau$ decay lifetime and universality are relevant; however, direct searches targeting $H^{\pm\pm} \to \ell^\pm \ell^\pm$ provide stronger limits.
\item {\it Electroweak Precision Data (EWPD):} The new Higgs bosons contribute to gauge boson self-energies, affecting the EWPD. These effects are well captured by the oblique parameters $S$, $T$, and $U$ at the one-loop level~\cite{Lavoura:1993nq, Kanemura:2012rs}; see Supplemental Material 
for their expressions. 
Notably, the triplet Higgs bosons also modify the $T$ parameter at tree level (or, rather, the $\rho$ parameter): $\alpha_{\rm em} T_{\rm tree} \equiv \rho-1 = -2v_\Delta^2/v^2$,  where  $\alpha_{\rm em}^{}=e^2/4\pi$ is the electromagnetic fine-structure constant. This gives an {\it upper} bound on $v_\Delta \lesssim \mathcal{O}(1)$~GeV, which justifies the limit $v_\Delta\ll v$ used in  Eq.~\eqref{eq:sum}. 

In Fig.~\ref{fig:EWPD}, we present the region of the model parameter space consistent with the EWPD at 68\% and 95\% confidence level (CL) in the $m_{H^{\pm\pm}}$--$\Delta m$ plane. Here, we adopt the EWPD fit results from PDG 2024~\cite{ParticleDataGroup:2024cfk}.\footnote{Currently, no global fit incorporates the latest ATLAS~\cite{ATLAS:2024erm} and CMS~\cite{CMS:2024lrd} measurements of the $W$-boson mass. However, as these results are fully compatible with the PDG 2024 combination~\cite{ParticleDataGroup:2024cfk}, we expect that their inclusion would lead to only a marginal shift in the best-fit values of $S$, $T$, and $U$, leaving the overall phenomenology largely unchanged. Further, in this work, we do not include the latest CDF result for the $W$-mass measurement~\cite{CDF:2022hxs}---which would have favored a larger $\Delta m$~\cite{Heeck:2022fvl,Bahl:2022gqg,Cheng:2022hbo, Butterworth:2022dkt}---since it has not been confirmed by the most recent measurements from ATLAS~\cite{ATLAS:2024erm} and CMS~\cite{CMS:2024lrd}.} We see that a mass splitting up to about 30 (40) GeV is allowed for $\Delta m>(<)0$ by the EWPD. The corresponding excluded region is indicated by the yellow-shaded area in Fig.~\ref{fig:decay}.

\begin{figure}[t!]
\centering
\includegraphics[width=0.42\textwidth]{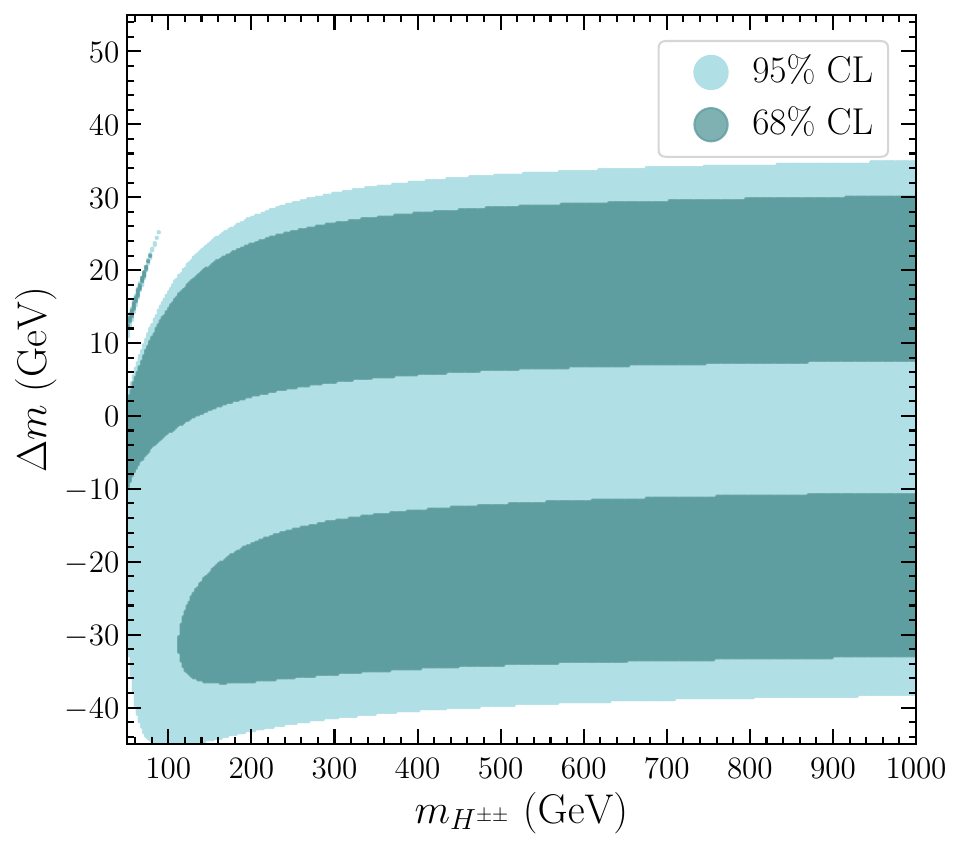}
\caption{Region of the parameter space allowed by the EWPD (taken from PDG~\cite{ParticleDataGroup:2024cfk}) at 68\% and 95\% CL in the $m_{H^{\pm\pm}}$--$\Delta m$ plane.}
\label{fig:EWPD}
\end{figure}
\item {\it SM Higgs Signal Strength:} The new Higgs bosons can significantly affect precision observables of the SM Higgs, particularly the global and diphoton signal strengths (the discussion of the latter is deferred to the next section). To a very good approximation, the global signal strength is just scaled by $\cos^2\alpha$. 
By combining the Tevatron and LHC measurements, Ref.~\cite{Heo:2024cif} estimates this value to be $\mu_{h, \rm {global}} = 1.012 \pm 0.034.$ This translates into a limit of $|\alpha| \lesssim 0.24\,(0.30) $ at $2\sigma \,(3\sigma)$.
\end{enumerate}

\section{Higgs Diphoton signal strength}
\label{sec:diphoton}
The inclusive diphoton signal strength of the SM Higgs boson is defined as
\begin{align}
\mu_{h \to \gamma\gamma} = \frac{\sigma_h}{\sigma_h^{\rm SM}} \times \frac{\Gamma_{h \to \gamma\gamma}/\Gamma_{h, {\rm tot}}}{\Gamma_{h \to \gamma\gamma}^{\rm SM}/\Gamma_{h, {\rm tot}}^{\rm SM}} \; ,
\label{eq:mugg}
\end{align}
where $\sigma_h/\sigma_h^{\rm SM} \approx \cos^2\alpha$ is the production cross section of $h$ normalized to its SM value, and $\Gamma_{h, {\rm tot}}$ is the total decay width of $h$ with its SM value $\Gamma_{h, {\rm tot}}^{\rm SM} = (4.088\pm 0.014)$~MeV at $m_h=125~{\rm GeV}$~\cite{LHCHiggsCrossSectionWorkingGroup:2016ypw}. The loop-induced diphoton decay rate is given by~\cite{Shifman:1979eb, Chun:2012jw, Dev:2013ff}
\begin{align}
&\Gamma^{(\rm SM)}_{h \to \gamma\gamma} = \frac{\alpha_{\rm em}^2 G_F m_h^3}{128\sqrt{2}\pi^3} \left|g^{(\rm SM)}_{h\gamma\gamma}\right|^2 \; ,
\label{eq:gamma}
\end{align}
where $G_F$ is the Fermi coupling, $\alpha_{\rm em}$ should be taken at the scale $q^2 = 0$ since the final state photons are real, and the effective couplings $g_{h\gamma\gamma}^{\rm (SM)}$ are given in the Supplemental Material. 

To obtain the total decay width $\Gamma_{h, {\rm tot}}$, one needs to compute all other decays of $h$ in the present model. The decay rates for $f\bar f/gg, WW$, and $ZZ$ modes can be obtained by scaling the corresponding SM decay rates with $(g^h_{ff})^2$, $(g^h_{WW})^2$, and $(g^h_{ZZ})^2$. These couplings, normalized to their SM value, are given by
\begin{align}
&g_{ff}^h \approx \cos\alpha \; ,
\\
&g_{WW}^h \approx \cos\alpha + \frac{2v_\Delta}{v} \sin\alpha \; ,
\\
&g_{ZZ}^h \approx \cos\alpha + \frac{4v_\Delta}{v} \sin\alpha \; .
\end{align}

The loop-induced $h\to Z\gamma$ decay also gets additional contributions from the triplet scalars~\cite{Dev:2013ff, 
Chen:2013vi, Chen:2013dh} (see Supplemental Material) and this is properly taken into account in our numerical calculation. 
The current experimental uncertainty in the $h\to Z\gamma$ channel is $\sim 70\%$~\cite{ATLAS:2023yqk}, compared to the $\sim 8\%$ precision for $h \to \gamma\gamma$, making the diphoton channel the most powerful indirect probe at present. Future improvements in $h \to Z\gamma$ could provide a complementary handle. 
%

\section{Precision Diphoton probe of the elusive parameter space}
\label{sec:analysis}
\begin{figure*}[htb!]
\centering
\includegraphics[width=0.98\textwidth]{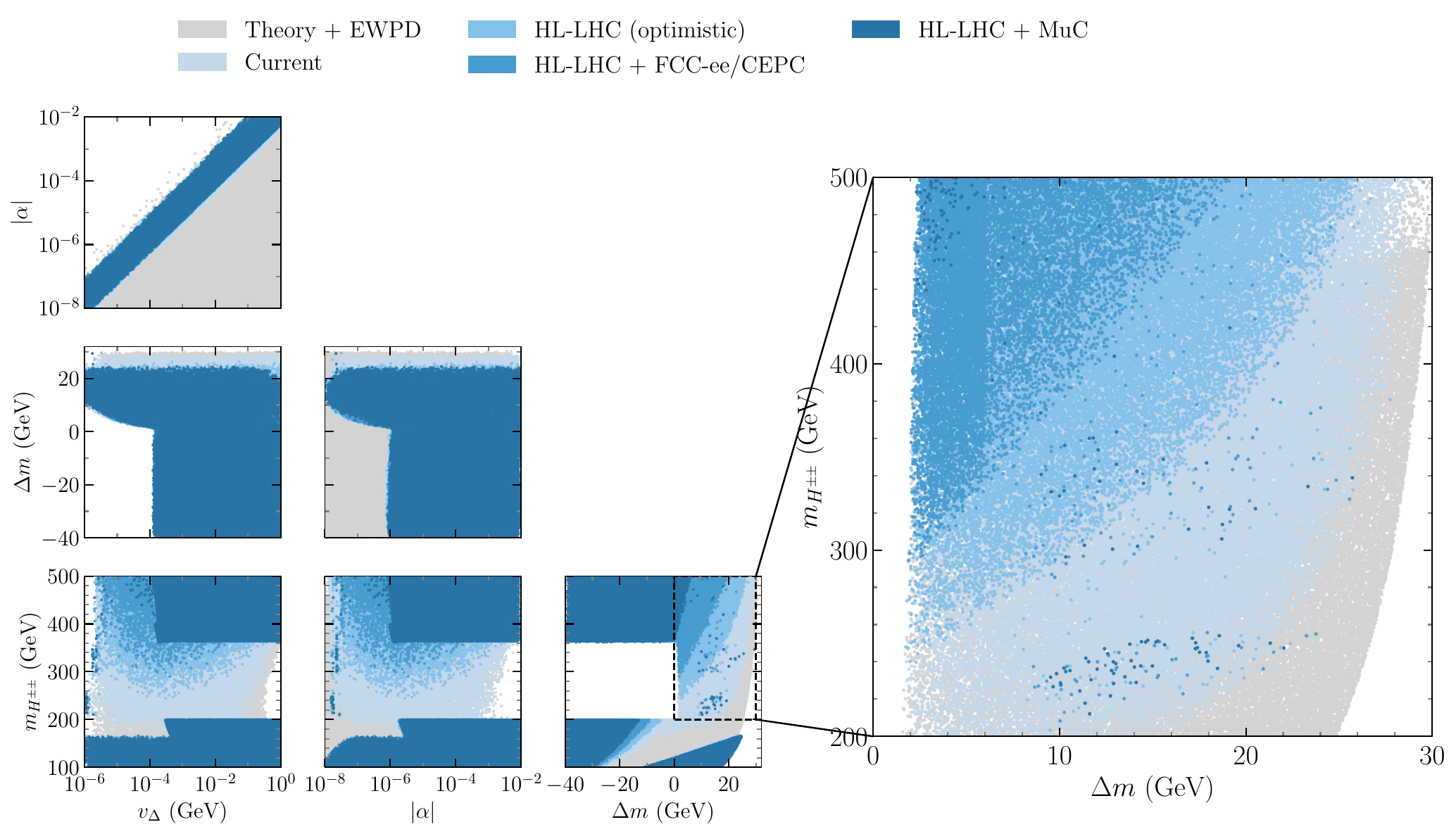}
\caption{Direct-search-inaccessible parameter space allowed by theoretical constraints and EWPD at 95\% CL (light gray). Overlaid regions in blue shades indicate the parameter space compatible with future $\mu_{h\to\gamma\gamma}$ measurements of increasing precision (darker blue represents higher precision). The enlarged panel on the right highlights the cascade-decay regime, characterized by the sequential decay $H^{\pm\pm} \to H^{\pm} \to H^0/A^0$, which can be almost completely probed by the expected precision to be achieved at a future MuC.}
\label{fig:corner}
\end{figure*}

As noted earlier, the triplet Higgs parameter space where cascade decays of $H^{\pm\pm}$ and $H^\pm$ into $H^0/A^0$ and off-shell $W$ bosons are dominant (illustrated in blue in Fig.~\ref{fig:decay}) remains beyond the reach of current direct search strategies. Also, the region with low-mass $H^{\pm\pm}$ (84--200 GeV) decaying into $W^\pm W^\pm$ is yet to be excluded~\cite{Kanemura:2014goa, Ashanujjaman:2022ofg}. Figure~\ref{fig:corner} shows this direct-search-inaccessible parameter space currently {\it allowed} by theoretical constraints and EWPD at 95\% CL, depicted as the light gray region in the background. We now assess the sensitivity of precision SM Higgs diphoton measurements to probe this region.

The hadron collider experiments at the Tevatron ($\sqrt s=1.96$ TeV) and the LHC ($\sqrt s=7$, 8, and 13 TeV) have performed several measurements of the inclusive diphoton signal strength of the SM Higgs boson. A recent combined analysis of these measurements, presented in Ref.~\cite{Heo:2024cif}, reports a signal strength of $0.99 \pm 0.04$ for the combined Higgs production and $1.10 \pm 0.07$ for the combined $h \to \gamma\gamma$ decay. Consequently, from Eq.~\eqref{eq:mugg}, the combined $\mu_{h \to \gamma\gamma}$ is
\begin{align}
\mu_{h \to \gamma\gamma} = 1.09 \pm 0.08 \; .
\label{eq:global}
\end{align}
Here we analyze four benchmark scenarios reflecting different levels of precision in measuring $\mu_{h\to\gamma\gamma}$. The first scenario corresponds to the current global average~\eqref{eq:global}, while the remaining three represent anticipated improvements from future measurements/experiments:
\begin{enumerate}[leftmargin=1.35em] 
\item Current: $\mu_{h \to \gamma\gamma} = 1.09 \pm 0.08$~\cite{Heo:2024cif},  
\item HL-LHC (optimistic): $\mu_{h \to \gamma\gamma} = 1.00 \pm 0.019$~\cite{deBlas:2019rxi},
\item HL-LHC+FCC-ee/CEPC: $\mu_{h \to \gamma\gamma} = 1.00 \pm 0.013$~\cite{deBlas:2019rxi},
\item HL-LHC+10 TeV MuC: $\mu_{h \to \gamma\gamma} = 1.00 \pm 0.007$~\cite{Castelli:2025mqk}.
\end{enumerate}
Note that the central value for all future measurements is assumed to be consistent with the SM (which serves as a good reference point), while the precision of the future measurements is reflected in the error bar. We emphasize that the results presented in this work are driven primarily by the anticipated improvement in experimental precision, and are therefore robust against modest shifts in the assumed central value of $\mu_{h \to \gamma\gamma}$.

The resulting allowed regions in the model parameter space are shown in Fig.~\ref{fig:corner}, where the model parameters are varied in the ranges shown. The white regions are already ruled out. The currently allowed region is shown in light blue, while projected future sensitivities are illustrated with progressively darker shades of blue: light blue (HL-LHC optimistic), normal blue (HL-LHC + FCC-ee/CEPC), and dark blue (HL-LHC + 10 TeV MuC). The present constraint---driven by the $\approx 8\%$ experimental uncertainty---still permits a wide swath of parameter space that is inaccessible to direct searches, particularly in the cascade-dominated region characterized by the sequential decay $H^{\pm\pm} \to H^{\pm} \to H^0/A^0$ (see the enlarged panel of Fig.~\ref{fig:corner}). However, the progressively darker shaded regions illustrate how rapidly this region shrinks once percent-level precision is achieved. Note also that throughout the allowed region one finds $|\alpha| \sim v_\Delta / v_\Phi$, i.e., $\alpha$ is typically two orders of magnitude smaller than the triplet VEV, reflecting the expected suppression of doublet-triplet mixing.



The \text{HL-LHC (optimistic)} scenario already removes a substantial portion of the parameter space, especially for larger $|\Delta m|$ or lighter $H^{\pm\pm}$, where the charged-scalar loop contributions tend to drive $\mu_{h\to\gamma\gamma}$ away from unity. Once subpercent precision is achieved---through the HL-LHC in combination with FCC-ee/CEPC or, in particular, with 10 TeV MuC---the surviving parameter space is further reduced. For the HL-LHC+FCC-ee/CEPC projection, the allowed region becomes noticeably narrower, corresponding mainly to heavy or nearly degenerate triplet spectra. In contrast, for the HL-LHC+10 TeV MuC case, the viable points become sparse, surviving only as a few isolated points scattered throughout the plane.


\section{Summary and Outlook}
\label{sec:summary}
We have examined the region of the type-II seesaw parameter space that remains unconstrained by current LHC searches. This ``elusive’’ region, consistent with perturbative unitarity, vacuum stability, and electroweak precision data, extends to light tripletlike Higgs bosons and allows sizable mass splittings among them.

We showed that precision measurements of the Higgs diphoton signal strength, $\mu_{h\to\gamma\gamma}$, which gets loop contributions from the charged scalars $H^\pm$ and $H^{\pm\pm}$, provides a sensitive indirect probe of this uncharted parameter space. Using the current global average and projected precisions from the HL-LHC, FCC-ee/CEPC, and 10 TeV MuC, we identified the parameter regions compatible with each scenario. While the present $\sim 8\%$ uncertainty leaves much of the direct-search-inaccessible space open, percent-level precision at the HL-LHC already removes a substantial portion of it. The subpercent level precision measurements anticipated at future lepton colliders further restrict the model: the HL-LHC+FCC-ee/CEPC projection favors relatively heavy or nearly degenerate triplet spectra, whereas the MuC leaves only a few isolated points surviving across the plane.

In summary, precision Higgs measurements offer a robust and complementary probe of the type-II seesaw mechanism. Future improvements in the Higgs-to-diphoton signal strength are poised to play a central role in extending the experimental sensitivity to the region inaccessible to direct searches.

\acknowledgments This research is supported by the Deutsche Forschungsgemeinschaft (DFG, German Research Foundation) under grant 396021762 - TRR 257, the National Natural Science Foundation of China under Grant No.~12475113, the CAS Project for Young Scientists in Basic Research (YSBR-099), and the Scientific and Technological Innovation Program of IHEP under Grant No.~E55457U2. The work of B.D. was partly supported
by the US Department of Energy under grant No. DE-SC0017987 and by a Humboldt
Fellowship from the Alexander von Humboldt Foundation. B.D. thanks the Theory Group at IHEP, Beijing for warm hospitality where this work was initiated, and also thanks the Astroparticle Group at KIT for a seminar visit where part of this work was done.

\bibliography{v0}

@article{Weinberg:1979sa,
    author = "Weinberg, Steven",
    title = "{Baryon and Lepton Nonconserving Processes}",
    reportNumber = "HUTP-79-A050",
    doi = "10.1103/PhysRevLett.43.1566",
    journal = "Phys. Rev. Lett.",
    volume = "43",
    pages = "1566--1570",
    year = "1979"
}

@article{Ma:1998dn,
    author = "Ma, Ernest",
    title = "{Pathways to naturally small neutrino masses}",
    eprint = "hep-ph/9805219",
    archivePrefix = "arXiv",
    reportNumber = "UCRHEP-T-222",
    doi = "10.1103/PhysRevLett.81.1171",
    journal = "Phys. Rev. Lett.",
    volume = "81",
    pages = "1171--1174",
    year = "1998"
}

@article{Konetschny:1977bn,
    author = "Konetschny, W. and Kummer, W.",
    title = "{Nonconservation of Total Lepton Number with Scalar Bosons}",
    reportNumber = "Print-77-0625 (VIENNA)",
    doi = "10.1016/0370-2693(77)90407-5",
    journal = "Phys. Lett. B",
    volume = "70",
    pages = "433--435",
    year = "1977"
}

@article{Cheng:1980qt,
    author = "Cheng, T. P. and Li, Ling-Fong",
    title = "{Neutrino Masses, Mixings and Oscillations in SU(2) x U(1) Models of Electroweak Interactions}",
    reportNumber = "PRINT-80-0511 (CARNEGIE-MELLON), COO-3066-152",
    doi = "10.1103/PhysRevD.22.2860",
    journal = "Phys. Rev. D",
    volume = "22",
    pages = "2860",
    year = "1980"
}

@article{Lazarides:1980nt,
    author = "Lazarides, George and Shafi, Q. and Wetterich, C.",
    title = "{Proton Lifetime and Fermion Masses in an SO(10) Model}",
    reportNumber = "FREIBURG-THEP-80-2",
    doi = "10.1016/0550-3213(81)90354-0",
    journal = "Nucl. Phys. B",
    volume = "181",
    pages = "287--300",
    year = "1981"
}

@article{Schechter:1980gr,
    author = "Schechter, J. and Valle, J. W. F.",
    title = "{Neutrino Masses in SU(2) x U(1) Theories}",
    reportNumber = "SU-4217-167, COO-3533-167",
    doi = "10.1103/PhysRevD.22.2227",
    journal = "Phys. Rev. D",
    volume = "22",
    pages = "2227",
    year = "1980"
}

@article{Mohapatra:1980yp,
    author = "Mohapatra, Rabindra N. and Senjanovic, Goran",
    title = "{Neutrino Masses and Mixings in Gauge Models with Spontaneous Parity Violation}",
    reportNumber = "FERMILAB-PUB-80-061-THY, FERMILAB-PUB-80-061-T",
    doi = "10.1103/PhysRevD.23.165",
    journal = "Phys. Rev. D",
    volume = "23",
    pages = "165",
    year = "1981"
}

@article{Magg:1980ut,
    author = "Magg, M. and Wetterich, C.",
    title = "{Neutrino Mass Problem and Gauge Hierarchy}",
    reportNumber = "CERN-TH-2829",
    doi = "10.1016/0370-2693(80)90825-4",
    journal = "Phys. Lett. B",
    volume = "94",
    pages = "61--64",
    year = "1980"
}

@article{Huitu:1996su,
    author = "Huitu, K. and Maalampi, J. and Pietila, A. and Raidal, M.",
    title = "{Doubly charged Higgs at LHC}",
    eprint = "hep-ph/9606311",
    archivePrefix = "arXiv",
    reportNumber = "HU-SEFT-R-1996-16, FTUV-96-34A, IFIC-96-40, TURKU-SFL-R-14",
    doi = "10.1016/S0550-3213(97)87466-4",
    journal = "Nucl. Phys. B",
    volume = "487",
    pages = "27--42",
    year = "1997"
}

@article{Gunion:1996pq,
    author = "Gunion, J. F. and Loomis, C. and Pitts, K. T.",
    title = "{Searching for doubly charged Higgs bosons at future colliders}",
    eprint = "hep-ph/9610237",
    archivePrefix = "arXiv",
    reportNumber = "UCD-96-30, FERMILAB-CONF-96-347, CDF-PUB-EXOTIC-PUBLIC-3922, SNOWMASS-1996-LTH096",
    journal = "eConf",
    volume = "C960625",
    pages = "LTH096",
    year = "1996"
}

@article{Chakrabarti:1998qy,
    author = "Chakrabarti, Surajit and Choudhury, Debajyoti and Godbole, Rohini M. and Mukhopadhyaya, Biswarup",
    title = "{Observing doubly charged Higgs bosons in photon-photon collisions}",
    eprint = "hep-ph/9804297",
    archivePrefix = "arXiv",
    reportNumber = "MRI-PHY-P980342",
    doi = "10.1016/S0370-2693(98)00743-6",
    journal = "Phys. Lett. B",
    volume = "434",
    pages = "347--353",
    year = "1998"
}

@article{Muhlleitner:2003me,
    author = "Muhlleitner, Margarete and Spira, Michael",
    title = "{A Note on doubly charged Higgs pair production at hadron colliders}",
    eprint = "hep-ph/0305288",
    archivePrefix = "arXiv",
    reportNumber = "PSI-PR-03-07",
    doi = "10.1103/PhysRevD.68.117701",
    journal = "Phys. Rev. D",
    volume = "68",
    pages = "117701",
    year = "2003"
}

@article{Chun:2003ej,
    author = "Chun, Eung Jin and Lee, Kang Young and Park, Seong Chan",
    title = "{Testing Higgs triplet model and neutrino mass patterns}",
    eprint = "hep-ph/0304069",
    archivePrefix = "arXiv",
    reportNumber = "KIAS-P03023",
    doi = "10.1016/S0370-2693(03)00770-6",
    journal = "Phys. Lett. B",
    volume = "566",
    pages = "142--151",
    year = "2003"
}

@article{Kakizaki:2003jk,
    author = "Kakizaki, Mitsuru and Ogura, Yoshiteru and Shima, Fumitaka",
    title = "{Lepton flavor violation in the triplet Higgs model}",
    eprint = "hep-ph/0304254",
    archivePrefix = "arXiv",
    reportNumber = "TU-689",
    doi = "10.1016/S0370-2693(03)00833-5",
    journal = "Phys. Lett. B",
    volume = "566",
    pages = "210--216",
    year = "2003"
}

@article{Akeroyd:2005gt,
    author = "Akeroyd, A. G. and Aoki, Mayumi",
    title = "{Single and pair production of doubly charged Higgs bosons at hadron colliders}",
    eprint = "hep-ph/0506176",
    archivePrefix = "arXiv",
    reportNumber = "KEK-TH-1023",
    doi = "10.1103/PhysRevD.72.035011",
    journal = "Phys. Rev. D",
    volume = "72",
    pages = "035011",
    year = "2005"
}

@article{Akeroyd:2007zv,
    author = "Akeroyd, A. G. and Aoki, Mayumi and Sugiyama, Hiroaki",
    title = "{Probing Majorana Phases and Neutrino Mass Spectrum in the Higgs Triplet Model at the CERN LHC}",
    eprint = "0712.4019",
    archivePrefix = "arXiv",
    primaryClass = "hep-ph",
    reportNumber = "SISSA-99-2007-EP",
    doi = "10.1103/PhysRevD.77.075010",
    journal = "Phys. Rev. D",
    volume = "77",
    pages = "075010",
    year = "2008"
}

@article{Garayoa:2007fw,
    author = "Garayoa, Julia and Schwetz, Thomas",
    title = "{Neutrino mass hierarchy and Majorana CP phases within the Higgs triplet model at the LHC}",
    eprint = "0712.1453",
    archivePrefix = "arXiv",
    primaryClass = "hep-ph",
    reportNumber = "CERN-PH-TH-2007-255, IFIC-07-75, FTUV-07-1210",
    doi = "10.1088/1126-6708/2008/03/009",
    journal = "JHEP",
    volume = "03",
    pages = "009",
    year = "2008"
}

@article{Han:2007bk,
    author = "Han, Tao and Mukhopadhyaya, Biswarup and Si, Zongguo and Wang, Kai",
    title = "{Pair production of doubly-charged scalars: Neutrino mass constraints and signals at the LHC}",
    eprint = "0706.0441",
    archivePrefix = "arXiv",
    primaryClass = "hep-ph",
    reportNumber = "MADPH-07-1488, HRI-P-07-05-001",
    doi = "10.1103/PhysRevD.76.075013",
    journal = "Phys. Rev. D",
    volume = "76",
    pages = "075013",
    year = "2007"
}

@article{Kadastik:2007yd,
    author = "Kadastik, M. and Raidal, M. and Rebane, L.",
    title = "{Direct determination of neutrino mass parameters at future colliders}",
    eprint = "0712.3912",
    archivePrefix = "arXiv",
    primaryClass = "hep-ph",
    doi = "10.1103/PhysRevD.77.115023",
    journal = "Phys. Rev. D",
    volume = "77",
    pages = "115023",
    year = "2008"
}

@article{delAguila:2008cj,
    author = "del Aguila, F. and Aguilar-Saavedra, J. A.",
    title = "{Distinguishing seesaw models at LHC with multi-lepton signals}",
    eprint = "0808.2468",
    archivePrefix = "arXiv",
    primaryClass = "hep-ph",
    doi = "10.1016/j.nuclphysb.2008.12.029",
    journal = "Nucl. Phys. B",
    volume = "813",
    pages = "22--90",
    year = "2009"
}

@article{FileviezPerez:2008jbu,
    author = "Fileviez Perez, Pavel and Han, Tao and Huang, Gui-yu and Li, Tong and Wang, Kai",
    title = "{Neutrino Masses and the CERN LHC: Testing Type II Seesaw}",
    eprint = "0805.3536",
    archivePrefix = "arXiv",
    primaryClass = "hep-ph",
    reportNumber = "MADPH-08-1510, NSF-KITP-08-65",
    doi = "10.1103/PhysRevD.78.015018",
    journal = "Phys. Rev. D",
    volume = "78",
    pages = "015018",
    year = "2008"
}

@article{Akeroyd:2009hb,
    author = "Akeroyd, A. G. and Chiang, Cheng-Wei",
    title = "{Doubly charged Higgs bosons and three-lepton signatures in the Higgs Triplet Model}",
    eprint = "0909.4419",
    archivePrefix = "arXiv",
    primaryClass = "hep-ph",
    doi = "10.1103/PhysRevD.80.113010",
    journal = "Phys. Rev. D",
    volume = "80",
    pages = "113010",
    year = "2009"
}

@article{Akeroyd:2009nu,
    author = "Akeroyd, A. G. and Aoki, Mayumi and Sugiyama, Hiroaki",
    title = "{Lepton Flavour Violating Decays tau ---{\ensuremath{>}} anti-l ll and mu ---{\ensuremath{>}} e gamma in the Higgs Triplet Model}",
    eprint = "0904.3640",
    archivePrefix = "arXiv",
    primaryClass = "hep-ph",
    reportNumber = "TU-844, SISSA-21-2009-EP",
    doi = "10.1103/PhysRevD.79.113010",
    journal = "Phys. Rev. D",
    volume = "79",
    pages = "113010",
    year = "2009"
}

@article{Akeroyd:2010ip,
    author = "Akeroyd, A. G. and Chiang, Cheng-Wei and Gaur, Naveen",
    title = "{Leptonic signatures of doubly charged Higgs boson production at the LHC}",
    eprint = "1009.2780",
    archivePrefix = "arXiv",
    primaryClass = "hep-ph",
    reportNumber = "SHEP-10-29",
    doi = "10.1007/JHEP11(2010)005",
    journal = "JHEP",
    volume = "11",
    pages = "005",
    year = "2010"
}

@article{Rodejohann:2010bv,
    author = "Rodejohann, Werner and Zhang, He",
    title = "{Higgs triplets at like-sign linear colliders and neutrino mixing}",
    eprint = "1011.3606",
    archivePrefix = "arXiv",
    primaryClass = "hep-ph",
    doi = "10.1103/PhysRevD.83.073005",
    journal = "Phys. Rev. D",
    volume = "83",
    pages = "073005",
    year = "2011"
}

@article{Rodejohann:2010jh,
    author = "Rodejohann, Werner",
    title = "{Inverse Neutrino-less Double Beta Decay Revisited: Neutrinos, Higgs Triplets and a Muon Collider}",
    eprint = "1005.2854",
    archivePrefix = "arXiv",
    primaryClass = "hep-ph",
    doi = "10.1103/PhysRevD.81.114001",
    journal = "Phys. Rev. D",
    volume = "81",
    pages = "114001",
    year = "2010"
}

@article{Melfo:2011nx,
    author = "Melfo, Alejandra and Nemevsek, Miha and Nesti, Fabrizio and Senjanovic, Goran and Zhang, Yue",
    title = "{Type II Seesaw at LHC: The Roadmap}",
    eprint = "1108.4416",
    archivePrefix = "arXiv",
    primaryClass = "hep-ph",
    doi = "10.1103/PhysRevD.85.055018",
    journal = "Phys. Rev. D",
    volume = "85",
    pages = "055018",
    year = "2012"
}

@article{Aoki:2011pz,
    author = "Aoki, Mayumi and Kanemura, Shinya and Yagyu, Kei",
    title = "{Testing the Higgs triplet model with the mass difference at the LHC}",
    eprint = "1110.4625",
    archivePrefix = "arXiv",
    primaryClass = "hep-ph",
    reportNumber = "KANAZAWA-11-17, UT-HET-059",
    doi = "10.1103/PhysRevD.85.055007",
    journal = "Phys. Rev. D",
    volume = "85",
    pages = "055007",
    year = "2012"
}

@article{Akeroyd:2011zza,
    author = "Akeroyd, A. G. and Sugiyama, Hiroaki",
    title = "{Production of doubly charged scalars from the decay of singly charged scalars in the Higgs Triplet Model}",
    eprint = "1105.2209",
    archivePrefix = "arXiv",
    primaryClass = "hep-ph",
    reportNumber = "SHEP-11-09",
    doi = "10.1103/PhysRevD.84.035010",
    journal = "Phys. Rev. D",
    volume = "84",
    pages = "035010",
    year = "2011"
}

@article{Arhrib:2011uy,
    author = "Arhrib, A. and Benbrik, R. and Chabab, M. and Moultaka, G. and Peyranere, M. C. and Rahili, L. and Ramadan, J.",
    title = "{The Higgs Potential in the Type II Seesaw Model}",
    eprint = "1105.1925",
    archivePrefix = "arXiv",
    primaryClass = "hep-ph",
    doi = "10.1103/PhysRevD.84.095005",
    journal = "Phys. Rev. D",
    volume = "84",
    pages = "095005",
    year = "2011"
}

@article{Aoki:2012jj,
    author = "Aoki, Mayumi and Kanemura, Shinya and Kikuchi, Mariko and Yagyu, Kei",
    title = "{Radiative corrections to the Higgs boson couplings in the triplet model}",
    eprint = "1211.6029",
    archivePrefix = "arXiv",
    primaryClass = "hep-ph",
    doi = "10.1103/PhysRevD.87.015012",
    journal = "Phys. Rev. D",
    volume = "87",
    number = "1",
    pages = "015012",
    year = "2013"
}

@article{Arbabifar:2012bd,
    author = "Arbabifar, Fatemeh and Bahrami, Sahar and Frank, Mariana",
    title = "{Neutral Higgs Bosons in the Higgs Triplet Model with nontrivial mixing}",
    eprint = "1211.6797",
    archivePrefix = "arXiv",
    primaryClass = "hep-ph",
    reportNumber = "CUMQ-HEP-168",
    doi = "10.1103/PhysRevD.87.015020",
    journal = "Phys. Rev. D",
    volume = "87",
    number = "1",
    pages = "015020",
    year = "2013"
}

@article{Chiang:2012dk,
    author = "Chiang, Cheng-Wei and Nomura, Takaaki and Tsumura, Koji",
    title = "{Search for doubly charged Higgs bosons using the same-sign diboson mode at the LHC}",
    eprint = "1202.2014",
    archivePrefix = "arXiv",
    primaryClass = "hep-ph",
    doi = "10.1103/PhysRevD.85.095023",
    journal = "Phys. Rev. D",
    volume = "85",
    pages = "095023",
    year = "2012"
}

@article{Akeroyd:2012nd,
    author = "Akeroyd, A. G. and Moretti, S. and Sugiyama, Hiroaki",
    title = "{Five-lepton and six-lepton signatures from production of neutral triplet scalars in the Higgs Triplet Model}",
    eprint = "1201.5047",
    archivePrefix = "arXiv",
    primaryClass = "hep-ph",
    reportNumber = "SHEP-11-32",
    doi = "10.1103/PhysRevD.85.055026",
    journal = "Phys. Rev. D",
    volume = "85",
    pages = "055026",
    year = "2012"
}

@article{Chun:2012zu,
    author = "Chun, Eung Jin and Sharma, Pankaj",
    title = "{Same-Sign Tetra-Leptons from Type II Seesaw}",
    eprint = "1206.6278",
    archivePrefix = "arXiv",
    primaryClass = "hep-ph",
    reportNumber = "KIAS-P12035",
    doi = "10.1007/JHEP08(2012)162",
    journal = "JHEP",
    volume = "08",
    pages = "162",
    year = "2012"
}

@article{Chun:2012jw,
    author = "Chun, Eung Jin and Lee, Hyun Min and Sharma, Pankaj",
    title = "{Vacuum Stability, Perturbativity, EWPD and Higgs-to-diphoton rate in Type II Seesaw Models}",
    eprint = "1209.1303",
    archivePrefix = "arXiv",
    primaryClass = "hep-ph",
    reportNumber = "KIAS-P12055",
    doi = "10.1007/JHEP11(2012)106",
    journal = "JHEP",
    volume = "11",
    pages = "106",
    year = "2012"
}

@article{Dinh:2012bp,
    author = "Dinh, D. N. and Ibarra, A. and Molinaro, E. and Petcov, S. T.",
    title = "{The $\mu - e$ Conversion in Nuclei, $\mu \to e \gamma, \mu \to 3e$ Decays and TeV Scale See-Saw Scenarios of Neutrino Mass Generation}",
    eprint = "1205.4671",
    archivePrefix = "arXiv",
    primaryClass = "hep-ph",
    reportNumber = "FLAVOUR(267104)-ERC-15, SISSA-10-2012-EP, TUM-HEP-837-12, CFTP-12-007",
    doi = "10.1007/JHEP08(2012)125",
    journal = "JHEP",
    volume = "08",
    pages = "125",
    year = "2012",
    note = "[Erratum: JHEP 09, 023 (2013)]"
}

@article{Barrie:2022ake,
    author = "Barrie, N. D. and Petcov, S. T.",
    title = "{Lepton Flavour Violation tests of Type II Seesaw Leptogenesis}",
    eprint = "2210.02110",
    archivePrefix = "arXiv",
    primaryClass = "hep-ph",
    reportNumber = "CTPU-PTC-22-18, SISSA 17/2022/FSI",
    doi = "10.1007/JHEP01(2023)001",
    journal = "JHEP",
    volume = "01",
    pages = "001",
    year = "2023"
}

@article{delAguila:2013mia,
    author = "del {\'A}guila, Francisco and Chala, Mikael",
    title = "{LHC bounds on Lepton Number Violation mediated by doubly and singly-charged scalars}",
    eprint = "1311.1510",
    archivePrefix = "arXiv",
    primaryClass = "hep-ph",
    doi = "10.1007/JHEP03(2014)027",
    journal = "JHEP",
    volume = "03",
    pages = "027",
    year = "2014"
}

@article{Chun:2013vma,
    author = "Chun, Eung Jin and Sharma, Pankaj",
    title = "{Search for a doubly-charged boson in four lepton final states in type II seesaw}",
    eprint = "1309.6888",
    archivePrefix = "arXiv",
    primaryClass = "hep-ph",
    reportNumber = "KIAS-P13054",
    doi = "10.1016/j.physletb.2013.11.056",
    journal = "Phys. Lett. B",
    volume = "728",
    pages = "256--261",
    year = "2014"
}

@article{Kanemura:2013vxa,
    author = "Kanemura, Shinya and Yagyu, Kei and Yokoya, Hiroshi",
    title = "{First constraint on the mass of doubly-charged Higgs bosons in the same-sign diboson decay scenario at the LHC}",
    eprint = "1305.2383",
    archivePrefix = "arXiv",
    primaryClass = "hep-ph",
    reportNumber = "UT-HET-080",
    doi = "10.1016/j.physletb.2013.08.054",
    journal = "Phys. Lett. B",
    volume = "726",
    pages = "316--319",
    year = "2013"
}

@article{Dev:2013ff,
    author = "Dev, P. S. Bhupal and Ghosh, Dilip Kumar and Okada, Nobuchika and Saha, Ipsita",
    title = "{125 GeV Higgs Boson and the Type-II Seesaw Model}",
    eprint = "1301.3453",
    archivePrefix = "arXiv",
    primaryClass = "hep-ph",
    reportNumber = "MAN-HEP-2012-021",
    doi = "10.1007/JHEP03(2013)150",
    journal = "JHEP",
    volume = "03",
    pages = "150",
    year = "2013",
    note = "[Erratum: JHEP 05, 049 (2013)]"
}

@article{Kanemura:2014goa,
    author = "Kanemura, Shinya and Kikuchi, Mariko and Yagyu, Kei and Yokoya, Hiroshi",
    title = "{Bounds on the mass of doubly-charged Higgs bosons in the same-sign diboson decay scenario}",
    eprint = "1407.6547",
    archivePrefix = "arXiv",
    primaryClass = "hep-ph",
    reportNumber = "UT-HET-091",
    doi = "10.1103/PhysRevD.90.115018",
    journal = "Phys. Rev. D",
    volume = "90",
    number = "11",
    pages = "115018",
    year = "2014"
}

@article{Kanemura:2014ipa,
    author = "Kanemura, Shinya and Kikuchi, Mariko and Yokoya, Hiroshi and Yagyu, Kei",
    title = "{LHC Run-I constraint on the mass of doubly charged Higgs bosons in the same-sign diboson decay scenario}",
    eprint = "1412.7603",
    archivePrefix = "arXiv",
    primaryClass = "hep-ph",
    reportNumber = "UT-HET-098",
    doi = "10.1093/ptep/ptv071",
    journal = "PTEP",
    volume = "2015",
    pages = "051B02",
    year = "2015"
}

@article{kang:2014jia,
    author = "Kang, Zhaofeng and Li, Jinmian and Li, Tianjun and Liu, Yandong and Ning, Guo-Zhu",
    title = "{Light Doubly Charged Higgs Boson via the $WW^*$ Channel at LHC}",
    eprint = "1404.5207",
    archivePrefix = "arXiv",
    primaryClass = "hep-ph",
    doi = "10.1140/epjc/s10052-015-3774-1",
    journal = "Eur. Phys. J. C",
    volume = "75",
    number = "12",
    pages = "574",
    year = "2015"
}

@article{Han:2015hba,
    author = "Han, Zhi-Long and Ding, Ran and Liao, Yi",
    title = "{LHC Phenomenology of Type II Seesaw: Nondegenerate Case}",
    eprint = "1502.05242",
    archivePrefix = "arXiv",
    primaryClass = "hep-ph",
    doi = "10.1103/PhysRevD.91.093006",
    journal = "Phys. Rev. D",
    volume = "91",
    pages = "093006",
    year = "2015"
}

@article{Han:2015sca,
    author = "Han, Zhi-Long and Ding, Ran and Liao, Yi",
    title = "{LHC phenomenology of the type II seesaw mechanism: Observability of neutral scalars in the nondegenerate case}",
    eprint = "1506.08996",
    archivePrefix = "arXiv",
    primaryClass = "hep-ph",
    doi = "10.1103/PhysRevD.92.033014",
    journal = "Phys. Rev. D",
    volume = "92",
    number = "3",
    pages = "033014",
    year = "2015"
}

@article{Deppisch:2015qwa,
    author = "Deppisch, Frank F. and Dev, P. S. Bhupal  and Pilaftsis, Apostolos",
    title = "{Neutrinos and Collider Physics}",
    eprint = "1502.06541",
    archivePrefix = "arXiv",
    primaryClass = "hep-ph",
    reportNumber = "MAN-HEP-2014-15",
    doi = "10.1088/1367-2630/17/7/075019",
    journal = "New J. Phys.",
    volume = "17",
    number = "7",
    pages = "075019",
    year = "2015"
}

@article{Mitra:2016wpr,
    author = "Mitra, Manimala and Niyogi, Saurabh and Spannowsky, Michael",
    title = "{Type-II Seesaw Model and Multilepton Signatures at Hadron Colliders}",
    eprint = "1611.09594",
    archivePrefix = "arXiv",
    primaryClass = "hep-ph",
    doi = "10.1103/PhysRevD.95.035042",
    journal = "Phys. Rev. D",
    volume = "95",
    number = "3",
    pages = "035042",
    year = "2017"
}

@article{Blunier:2016peh,
    author = "Blunier, Sylvain and Cottin, Giovanna and D{\'\i}az, Marco Aurelio and Koch, Benjamin",
    title = "{Phenomenology of a Higgs triplet model at future $e^{+}e^{-}$ colliders}",
    eprint = "1611.07896",
    archivePrefix = "arXiv",
    primaryClass = "hep-ph",
    reportNumber = "CAVENDISH-HEP-16-19",
    doi = "10.1103/PhysRevD.95.075038",
    journal = "Phys. Rev. D",
    volume = "95",
    number = "7",
    pages = "075038",
    year = "2017"
}

@article{Das:2016bir,
    author = "Das, Dipankar and Santamaria, Arcadi",
    title = "{Updated scalar sector constraints in the Higgs triplet model}",
    eprint = "1604.08099",
    archivePrefix = "arXiv",
    primaryClass = "hep-ph",
    reportNumber = "FTUV-16-0426, IFIC-16-14",
    doi = "10.1103/PhysRevD.94.015015",
    journal = "Phys. Rev. D",
    volume = "94",
    number = "1",
    pages = "015015",
    year = "2016"
}

@article{Dev:2017ouk,
    author = "Dev, P. S. Bhupal and Vila, Clara Miralles and Rodejohann, Werner",
    title = "{Naturalness in testable type II seesaw scenarios}",
    eprint = "1703.00828",
    archivePrefix = "arXiv",
    primaryClass = "hep-ph",
    doi = "10.1016/j.nuclphysb.2017.06.007",
    journal = "Nucl. Phys. B",
    volume = "921",
    pages = "436--453",
    year = "2017"
}

@article{Ghosh:2017pxl,
    author = "Ghosh, Dilip Kumar and Ghosh, Nivedita and Saha, Ipsita and Shaw, Avirup",
    title = "{Revisiting the high-scale validity of the type II seesaw model with novel LHC signature}",
    eprint = "1711.06062",
    archivePrefix = "arXiv",
    primaryClass = "hep-ph",
    doi = "10.1103/PhysRevD.97.115022",
    journal = "Phys. Rev. D",
    volume = "97",
    number = "11",
    pages = "115022",
    year = "2018"
}

@article{Nomura:2017abh,
    author = "Nomura, Takaaki and Okada, Hiroshi and Yokoya, Hiroshi",
    title = "{Discriminating leptonic Yukawa interactions with doubly charged scalar at the ILC}",
    eprint = "1702.03396",
    archivePrefix = "arXiv",
    primaryClass = "hep-ph",
    reportNumber = "KIAS-P17012",
    doi = "10.1016/j.nuclphysb.2018.02.011",
    journal = "Nucl. Phys. B",
    volume = "929",
    pages = "193--206",
    year = "2018"
}

@article{Cai:2017mow,
    author = "Cai, Yi and Han, Tao and Li, Tong and Ruiz, Richard",
    title = "{Lepton Number Violation: Seesaw Models and Their Collider Tests}",
    eprint = "1711.02180",
    archivePrefix = "arXiv",
    primaryClass = "hep-ph",
    reportNumber = "PITT-PACC-1712, IPPP-17-74, COEPP-MN-17-17",
    doi = "10.3389/fphy.2018.00040",
    journal = "Front. in Phys.",
    volume = "6",
    pages = "40",
    year = "2018"
}

@article{Antusch:2018svb,
    author = "Antusch, Stefan and Fischer, Oliver and Hammad, A. and Scherb, Christiane",
    title = "{Low scale type II seesaw: Present constraints and prospects for displaced vertex searches}",
    eprint = "1811.03476",
    archivePrefix = "arXiv",
    primaryClass = "hep-ph",
    doi = "10.1007/JHEP02(2019)157",
    journal = "JHEP",
    volume = "02",
    pages = "157",
    year = "2019"
}

@article{Xu:2023ene,
    author = "Xu, Fang",
    title = "{Neutral and doubly charged scalars at future lepton colliders}",
    eprint = "2302.08653",
    archivePrefix = "arXiv",
    primaryClass = "hep-ph",
    doi = "10.1103/PhysRevD.108.036002",
    journal = "Phys. Rev. D",
    volume = "108",
    number = "3",
    pages = "036002",
    year = "2023"
}

@article{Jia:2024wqi,
    author = "Jia, Jie-Cheng and Han, Zhi-Long and Huang, Fei and Jin, Yi and Li, Honglei",
    title = "{Production of single doubly charged Higgs bosons at muon colliders}",
    eprint = "2409.16582",
    archivePrefix = "arXiv",
    primaryClass = "hep-ph",
    doi = "10.1103/PhysRevD.111.015009",
    journal = "Phys. Rev. D",
    volume = "111",
    number = "1",
    pages = "015009",
    year = "2025"
}

@article{Dev:2023nha,
    author = "Dev, P. S. Bhupal and Heeck, Julian and Thapa, Anil",
    title = "{Neutrino mass models at $\mu $TRISTAN}",
    eprint = "2309.06463",
    archivePrefix = "arXiv",
    primaryClass = "hep-ph",
    reportNumber = "CETUP-2023-005",
    doi = "10.1140/epjc/s10052-024-12496-0",
    journal = "Eur. Phys. J. C",
    volume = "84",
    number = "2",
    pages = "148",
    year = "2024"
}

@article{Lichtenstein:2023iut,
    author = "Lichtenstein, Gabriela and Schmidt, Michael A. and Valencia, German and Volkas, Raymond R.",
    title = "{Complementarity of $\mu$TRISTAN and Belle II in searches for charged-lepton flavour violation}",
    eprint = "2307.11369",
    archivePrefix = "arXiv",
    primaryClass = "hep-ph",
    doi = "10.1016/j.physletb.2023.138144",
    journal = "Phys. Lett. B",
    volume = "845",
    pages = "138144",
    year = "2023"
}

@article{Babu:2016rcr,
    author = "Babu, K. S. and Jana, Sudip",
    title = "{Probing Doubly Charged Higgs Bosons at the LHC through Photon Initiated Processes}",
    eprint = "1612.09224",
    archivePrefix = "arXiv",
    primaryClass = "hep-ph",
    reportNumber = "OSU-HEP-16-11",
    doi = "10.1103/PhysRevD.95.055020",
    journal = "Phys. Rev. D",
    volume = "95",
    number = "5",
    pages = "055020",
    year = "2017"
}

@article{Fuks:2019clu,
    author = "Fuks, Benjamin and Nemev{\v{s}}ek, Miha and Ruiz, Richard",
    title = "{Doubly Charged Higgs Boson Production at Hadron Colliders}",
    eprint = "1912.08975",
    archivePrefix = "arXiv",
    primaryClass = "hep-ph",
    reportNumber = "CP3-19-61, MCnet-19-29, VBSCAN-PUB-11-19",
    doi = "10.1103/PhysRevD.101.075022",
    journal = "Phys. Rev. D",
    volume = "101",
    number = "7",
    pages = "075022",
    year = "2020"
}

@article{Dev:2018vpr,
    author = "Dev, P. S. Bhupal and Mohapatra, Rabindra N. and Zhang, Yongchao",
    title = "{Probing TeV scale origin of neutrino mass at future lepton colliders via neutral and doubly-charged scalars}",
    eprint = "1803.11167",
    archivePrefix = "arXiv",
    primaryClass = "hep-ph",
    doi = "10.1103/PhysRevD.98.075028",
    journal = "Phys. Rev. D",
    volume = "98",
    number = "7",
    pages = "075028",
    year = "2018"
}

@article{Dev:2018tox,
    author = "Dev, P. S. Bhupal and Zhang, Yongchao",
    title = "{Displaced vertex signatures of doubly charged scalars in the type-II seesaw and its left-right extensions}",
    eprint = "1808.00943",
    archivePrefix = "arXiv",
    primaryClass = "hep-ph",
    doi = "10.1007/JHEP10(2018)199",
    journal = "JHEP",
    volume = "10",
    pages = "199",
    year = "2018"
}

@article{Crivellin:2018ahj,
    author = "Crivellin, Andreas and Ghezzi, Margherita and Panizzi, Luca and Pruna, Giovanni Marco and Signer, Adrian",
    title = "{Low- and high-energy phenomenology of a doubly charged scalar}",
    eprint = "1807.10224",
    archivePrefix = "arXiv",
    primaryClass = "hep-ph",
    reportNumber = "PSI-PR-18-08, ZU-TH-28/18",
    doi = "10.1103/PhysRevD.99.035004",
    journal = "Phys. Rev. D",
    volume = "99",
    number = "3",
    pages = "035004",
    year = "2019"
}

@article{Agrawal:2018pci,
    author = "Agrawal, Pankaj and Mitra, Manimala and Niyogi, Saurabh and Shil, Sujay and Spannowsky, Michael",
    title = "{Probing the Type-II Seesaw Mechanism through the Production of Higgs Bosons at a Lepton Collider}",
    eprint = "1803.00677",
    archivePrefix = "arXiv",
    primaryClass = "hep-ph",
    reportNumber = "IP/BBSR/2018-18, IPPP/18/14, IP-BBSR-2018-18, IPPP-18-14",
    doi = "10.1103/PhysRevD.98.015024",
    journal = "Phys. Rev. D",
    volume = "98",
    number = "1",
    pages = "015024",
    year = "2018"
}

@article{Dev:2018sel,
    author = "Dev, P. S. Bhupal and Ramsey-Musolf, Michael J. and Zhang, Yongchao",
    title = "{Doubly-Charged Scalars in the Type-II Seesaw Mechanism: Fundamental Symmetry Tests and High-Energy Searches}",
    eprint = "1806.08499",
    archivePrefix = "arXiv",
    primaryClass = "hep-ph",
    reportNumber = "ACFI T18-10, ACFI-T18-10",
    doi = "10.1103/PhysRevD.98.055013",
    journal = "Phys. Rev. D",
    volume = "98",
    number = "5",
    pages = "055013",
    year = "2018"
}

@article{deMelo:2019asm,
    author = "de Melo, Tessio B. and Queiroz, Farinaldo S. and Villamizar, Yoxara",
    title = "{Doubly Charged Scalar at the High-Luminosity and High-Energy LHC}",
    eprint = "1909.07429",
    archivePrefix = "arXiv",
    primaryClass = "hep-ph",
    reportNumber = "IIPDM-2019",
    doi = "10.1142/S0217751X19501574",
    journal = "Int. J. Mod. Phys. A",
    volume = "34",
    number = "27",
    pages = "1950157",
    year = "2019"
}

@article{Primulando:2019evb,
    author = "Primulando, R. and Julio, J. and Uttayarat, P.",
    title = "{Scalar phenomenology in type-II seesaw model}",
    eprint = "1903.02493",
    archivePrefix = "arXiv",
    primaryClass = "hep-ph",
    doi = "10.1007/JHEP08(2019)024",
    journal = "JHEP",
    volume = "08",
    pages = "024",
    year = "2019"
}

@article{Chun:2019hce,
    author = "Chun, Eung Jin and Khan, Sarif and Mandal, Sanjoy and Mitra, Manimala and Shil, Sujay",
    title = "{Same-sign tetralepton signature at the Large Hadron Collider and a future $pp$ collider}",
    eprint = "1911.00971",
    archivePrefix = "arXiv",
    primaryClass = "hep-ph",
    doi = "10.1103/PhysRevD.101.075008",
    journal = "Phys. Rev. D",
    volume = "101",
    number = "7",
    pages = "075008",
    year = "2020"
}

@article{Padhan:2019jlc,
    author = "Padhan, Rojalin and Das, Debottam and Mitra, Manimala and Kumar Nayak, Aruna",
    title = "{Probing doubly and singly charged Higgs bosons at the $pp$ collider HE-LHC}",
    eprint = "1909.10495",
    archivePrefix = "arXiv",
    primaryClass = "hep-ph",
    doi = "10.1103/PhysRevD.101.075050",
    journal = "Phys. Rev. D",
    volume = "101",
    number = "7",
    pages = "075050",
    year = "2020"
}

@article{Rahili:2019ixf,
    author = "Rahili, L. and Arhrib, A. and Benbrik, R.",
    title = "{Associated production of SM Higgs with a photon in type-II seesaw models at the ILC}",
    eprint = "1909.07793",
    archivePrefix = "arXiv",
    primaryClass = "hep-ph",
    doi = "10.1140/epjc/s10052-019-7471-3",
    journal = "Eur. Phys. J. C",
    volume = "79",
    number = "11",
    pages = "940",
    year = "2019"
}

@article{Dev:2019hev,
    author = "Dev, P. S. Bhupal and Khan, Sarif and Mitra, Manimala and Rai, Santosh Kumar",
    title = "{Doubly-charged Higgs boson at a future electron-proton collider}",
    eprint = "1903.01431",
    archivePrefix = "arXiv",
    primaryClass = "hep-ph",
    reportNumber = "HRI-RECAPP-2019-001, IP/BBSR/2019-1",
    doi = "10.1103/PhysRevD.99.115015",
    journal = "Phys. Rev. D",
    volume = "99",
    number = "11",
    pages = "115015",
    year = "2019"
}

@article{Gluza:2020qrt,
    author = "Gluza, Janusz and Kordiaczynska, Magdalena and Srivastava, Tripurari",
    title = "{Discriminating the HTM and MLRSM models in collider studies via doubly charged Higgs boson pair production and the subsequent leptonic decays}",
    eprint = "2006.04610",
    archivePrefix = "arXiv",
    primaryClass = "hep-ph",
    doi = "10.1088/1674-1137/abfe51",
    journal = "Chin. Phys. C",
    volume = "45",
    number = "7",
    pages = "073113",
    year = "2021"
}

@article{Bandyopadhyay:2020mnp,
    author = "Bandyopadhyay, Priyotosh and Karan, Anirban and Sen, Chandrima",
    title = "{Discerning Signatures of Seesaw Models and Complementarity of Leptonic Colliders}",
journal = "",
    eprint = "2011.04191",
    archivePrefix = "arXiv",
    primaryClass = "hep-ph",
    reportNumber = "IITH-PH-0008/20",
    month = "11",
    year = "2020"
}

@article{Ashanujjaman:2021txz,
    author = "Ashanujjaman, Saiyad and Ghosh, Kirtiman",
    title = "{Revisiting type-II see-saw: present limits and future prospects at LHC}",
    eprint = "2108.10952",
    archivePrefix = "arXiv",
    primaryClass = "hep-ph",
    doi = "10.1007/JHEP03(2022)195",
    journal = "JHEP",
    volume = "03",
    pages = "195",
    year = "2022"
}

@article{Yang:2021skb,
    author = "Yang, Xing-Hua and Yang, Zhong-Juan",
    title = "{Doubly charged Higgs production at future $ep$ colliders}",
    eprint = "2103.11412",
    archivePrefix = "arXiv",
    primaryClass = "hep-ph",
    doi = "10.1088/1674-1137/ac581b",
    journal = "Chin. Phys. C",
    volume = "46",
    number = "6",
    pages = "063107",
    year = "2022"
}

@article{Ashanujjaman:2022ofg,
    author = "Ashanujjaman, Saiyad and Ghosh, Kirtiman and Sahu, Rameswar",
    title = "{Low-mass doubly charged Higgs bosons at the LHC}",
    eprint = "2211.00632",
    archivePrefix = "arXiv",
    primaryClass = "hep-ph",
    doi = "10.1103/PhysRevD.107.015018",
    journal = "Phys. Rev. D",
    volume = "107",
    number = "1",
    pages = "015018",
    year = "2023"
}

@article{Butterworth:2022dkt,
    author = "Butterworth, Jon and Heeck, Julian and Jeon, Si Hyun and Mattelaer, Olivier and Ruiz, Richard",
    title = "{Testing the scalar triplet solution to CDF{\textquoteright}s heavy W problem at the LHC}",
    eprint = "2210.13496",
    archivePrefix = "arXiv",
    primaryClass = "hep-ph",
    reportNumber = "IRMP-CP3-22-47, IFJPAN-IV-2022-15, MCNET-22",
    doi = "10.1103/PhysRevD.107.075020",
    journal = "Phys. Rev. D",
    volume = "107",
    number = "7",
    pages = "075020",
    year = "2023"
}

@article{Ashanujjaman:2022tdn,
    author = "Ashanujjaman, Saiyad and Ghosh, Kirtiman and Huitu, Katri",
    title = "{Type-II see-saw: searching the LHC elusive low-mass triplet-like Higgses at $e^-e^+$ colliders}",
    eprint = "2205.14983",
    archivePrefix = "arXiv",
    primaryClass = "hep-ph",
    doi = "10.1103/PhysRevD.106.075028",
    journal = "Phys. Rev. D",
    volume = "106",
    number = "7",
    pages = "075028",
    year = "2022"
}

@article{Ashanujjaman:2023tlj,
    author = "Ashanujjaman, Saiyad and Maharathy, Siddharth P.",
    title = "{Probing compressed mass spectra in the type-II seesaw model at the LHC}",
    eprint = "2305.06889",
    archivePrefix = "arXiv",
    primaryClass = "hep-ph",
    doi = "10.1103/PhysRevD.107.115026",
    journal = "Phys. Rev. D",
    volume = "107",
    number = "11",
    pages = "115026",
    year = "2023"
}

@article{Li:2023ksw,
    author = "Li, Tong and Yao, Chang-Yuan and Yuan, Man",
    title = "{Revealing the origin of neutrino masses through the Type II Seesaw mechanism at high-energy muon colliders}",
    eprint = "2301.07274",
    archivePrefix = "arXiv",
    primaryClass = "hep-ph",
    reportNumber = "DESY-23-006",
    doi = "10.1007/JHEP03(2023)137",
    journal = "JHEP",
    volume = "03",
    pages = "137",
    year = "2023"
}

@article{Maharathy:2023dtp,
    author = "Maharathy, Siddharth P. and Mitra, Manimala",
    title = "{Type-II see-saw at {\ensuremath{\mu}}+{\ensuremath{\mu}}{\ensuremath{-}} collider}",
    eprint = "2304.08732",
    archivePrefix = "arXiv",
    primaryClass = "hep-ph",
    doi = "10.1016/j.physletb.2023.138105",
    journal = "Phys. Lett. B",
    volume = "844",
    pages = "138105",
    year = "2023"
}

@article{Fridell:2023gjx,
    author = "Fridell, K{\r{a}}re and Kitano, Ryuichiro and Takai, Ryoto",
    title = "{Lepton flavor physics at {\ensuremath{\mu}}$^{+}${\ensuremath{\mu}}$^{+}$ colliders}",
    eprint = "2304.14020",
    archivePrefix = "arXiv",
    primaryClass = "hep-ph",
    reportNumber = "KEK-TH-2519",
    doi = "10.1007/JHEP06(2023)086",
    journal = "JHEP",
    volume = "06",
    pages = "086",
    year = "2023"
}

@article{OPAL:2001luy,
    author = "Abbiendi, G. and others",
    collaboration = "OPAL",
    title = "{Search for doubly charged Higgs bosons with the OPAL detector at LEP}",
    eprint = "hep-ex/0111059",
    archivePrefix = "arXiv",
    reportNumber = "CERN-EP-2001-082",
    doi = "10.1016/S0370-2693(01)01474-5",
    journal = "Phys. Lett. B",
    volume = "526",
    pages = "221--232",
    year = "2002"
}

@article{DELPHI:2002bkf,
    author = "Abdallah, J. and others",
    collaboration = "DELPHI",
    title = "{Search for doubly charged Higgs bosons at LEP-2}",
    eprint = "hep-ex/0303026",
    archivePrefix = "arXiv",
    reportNumber = "CERN-EP-2002-077",
    doi = "10.1016/S0370-2693(02)03125-8",
    journal = "Phys. Lett. B",
    volume = "552",
    pages = "127--137",
    year = "2003"
}

@inproceedings{LEPHiggsWorkingGroupforHiggsbosonsearches:2001ogs,
    collaboration = "LEP Higgs Working Group for Higgs boson searches, ALEPH, DELPHI, L3, OPAL",
    title = "{Search for charged Higgs bosons: Preliminary combined results using LEP data collected at energies up to 209-GeV}",
    booktitle = "{2001 Europhysics Conference on High Energy Physics}",
    eprint = "hep-ex/0107031",
    archivePrefix = "arXiv",
    reportNumber = "LHWG-NOTE-2001-05, ALEPH-2001-043, PHYSICS-2001-016, DELPHI-2001-115, CERN-L3-NOTE-2689, OPAL-TN-696",
    month = "7",
    year = "2001"
}

@article{Datta:1999nc,
    author = "Datta, Anindya and Raychaudhuri, Amitava",
    title = "{Mass bounds for triplet scalars of the left-right symmetric model and their future detection prospects}",
    eprint = "hep-ph/9905421",
    archivePrefix = "arXiv",
    reportNumber = "CUPP-99-3",
    doi = "10.1103/PhysRevD.62.055002",
    journal = "Phys. Rev. D",
    volume = "62",
    pages = "055002",
    year = "2000"
}

@article{ATLAS:2014kca,
    author = "Aad, Georges and others",
    collaboration = "ATLAS",
    title = "{Search for anomalous production of prompt same-sign lepton pairs and pair-produced doubly charged Higgs bosons with $ \sqrt{s}=8 $ TeV $pp$ collisions using the ATLAS detector}",
    eprint = "1412.0237",
    archivePrefix = "arXiv",
    primaryClass = "hep-ex",
    reportNumber = "CERN-PH-EP-2014-158",
    doi = "10.1007/JHEP03(2015)041",
    journal = "JHEP",
    volume = "03",
    pages = "041",
    year = "2015"
}

@misc{CMS:2017pet,
    collaboration = "CMS",
    title = "{A search for doubly-charged Higgs boson production in three and four lepton final states at $\sqrt{s}=13~\mathrm{TeV}$}",
    howpublished = "CMS-PAS-HIG-16-036",
    year = "2017"
}

\clearpage
\newpage
\onecolumngrid

\begin{center}
\textbf{\large Supplemental Material}
\end{center}

\setcounter{section}{0}
\renewcommand{\thesection}{\Alph{section}}
\renewcommand{\theequation}{\thesection\arabic{equation}}

\setcounter{equation}{0}
\setcounter{figure}{0}
\setcounter{table}{0}
\setcounter{page}{1}
\makeatletter
\renewcommand{\theequation}{S\arabic{equation}}
\renewcommand{\thefigure}{S\arabic{figure}}
\renewcommand{\thetable}{S\arabic{table}}
\renewcommand{\thepage}{S\arabic{page}}

\section{Expressions for the Oblique parameters $S$, $T$ and $U$}
\label{app:STU}
To a good approximation, neglecting the effects of scalar mixing, the oblique parameters $S$, $T$, and $U$ are given by~\cite{Lavoura:1993nq}:\footnote{In Eq.~(4.1) of Ref.~\cite{Chun:2012jw}, the second loop function in the square brackets $\xi \left(m_{T_3^{}}^2/m_W^2, m_{T_3^{}}^2/m_W^2\right)$, appearing in the expression for the parameter $U$, should be corrected to $\xi \left(m_{T_3}^2/m_W^2, m_{T_3-1}^2/m_W^2\right)$.} 
\begin{align}
S =& -\frac{1}{3\pi} \ln \frac{m_{H^{\pm\pm}}^2}{m_{H^0}^2} - \frac{2}{\pi} \left[ (1-2s_W^2)^2 \xi\left(\frac{m_{H^{\pm\pm}}^2}{m^2_{Z}}, \frac{m_{H^{\pm\pm}}^2}{m^2_{Z}}\right) + s_W^4 \xi\left(\frac{m_{H^\pm}^2}{m^2_{Z}}, \frac{m_{H^\pm}^2}{m^2_{Z}}\right) + \xi\left(\frac{m_{H^0}^2}{m^2_{Z}}, \frac{m_{H^0}^2}{m^2_{Z}}\right) \right] \; ,
\\
T =&~\frac{1}{8\pi c_W^2 s_W^2} \left[ \eta\left(\frac{m_{H^{\pm\pm}}^2}{m_Z^2}, \frac{m_{H^\pm}^2}{m_Z^2}\right) + \eta\left(\frac{m_{H^{\pm}}^2}{m_Z^2}, \frac{m_{H^0}^2}{m_Z^2}\right) \right] \; ,
\\
U =&~\frac{1}{6\pi} \ln \frac{m_{H^\pm}^4}{m_{H^{\pm\pm}}^2 m_{H^0}^2} + \frac{2}{\pi} \left[ (1-2s_W^2)^2 \xi\left(\frac{m_{H^{\pm\pm}}^2}{m_Z^2}, \frac{m_{H^{\pm\pm}}^2}{m_Z^2}\right) + s_W^4 \xi\left(\frac{m_{H^{\pm}}^2}{m_Z^2}, \frac{m_{H^{\pm}}^2}{m_Z^2}\right) + \xi \left(\frac{m_{H^0}^2}{m_Z^2} \;, \frac{m_{H^0}^2}{m_Z^2}\right) \right] \nonumber
\\
& - \frac{2}{\pi} \left[ \xi\left(\frac{m_{H^{\pm\pm}}^2}{m_W^2}, \frac{m_{H^\pm}^2}{m_W^2}\right) + \xi \left(\frac{m_{H^{\pm}}^2}{m_W^2},\frac{m_{H^0}^2}{m_W^2}\right) \right]  \;,
\end{align}
where the functions are defined as
\begin{align}
\xi(x,y) =&~\frac{4}{9} - \frac{5}{12}(x+y) + \frac{1}{6}(x-y)^2 + \frac{1}{4} \left[ x^2-y^2-\frac{1}{3}(x-y)^3-\frac{x^2+y^2}{x-y} \right] \ln \frac{x}{y} - \frac{1}{12} d(x,y)f(x,y) \; ,
\\
\eta(x,y) =&~ x + y - \frac{2xy}{x-y} \ln \frac{x}{y} \; ,
\\
d(x,y) =&~ -1+2(x+y)-(x-y)^2 \; ,
\\
f(x,y) =& \begin{cases} 
-2\sqrt{d(x,y)} \left\{ \arctan \left[\displaystyle \frac{x-y+1}{\sqrt{d(x,y)}}\right] - \arctan \left[\displaystyle \frac{x-y-1}{\sqrt{d(x,y)}} \right]\right\} \, & d(x,y) > 0 
\\
\sqrt{-d(x,y)}\ln \left[ \displaystyle\frac{x+y-1+\sqrt{-d(x,y)}}{x+y-1-\sqrt{-d(x,y)}} \right] \, & d(x,y) \le 0
\end{cases} \;.
\end{align}

\section{Relevant loop functions in the loop-induced Higgs decays $h\to\gamma\gamma$ and $h\to Z\gamma$}
\label{app:LoopFunc}
\subsection{$h\to \gamma\gamma$}
The loop-induced $h\to \gamma\gamma$ decay rate in the SM and in the type-II seesaw model are respectively given by 
\begin{align}
&\Gamma^{(\rm SM)}_{h \to \gamma\gamma} = \frac{\alpha_{\rm em}^2 G_F m_h^3}{128\sqrt{2}\pi^3} \left|g^{(\rm SM)}_{h\gamma\gamma}\right|^2 \; ,
\label{eq:gamma}
\end{align}
where the effective couplings $g_{h\gamma\gamma}^{(\rm SM)}$ 
are given by~\cite{Shifman:1979eb, Chun:2012jw, Dev:2013ff} 
\begin{align}
g_{h\gamma\gamma}^{\rm SM} =& \sum_f N^c_f Q_f^2 \beta^{1/2}_{\gamma\gamma}\left(r_f\right) + \beta^1_{\gamma\gamma}\left(r_W\right) \; ,
\\
g_{h\gamma\gamma} =& \sum_f N^c_f Q_f^2 g^h_{ff} \beta^{1/2}_{\gamma\gamma}\left(r_f\right) + g^h_{WW} \beta^1_{\gamma\gamma}\left(r_W\right) + \frac{\lambda_{hH^+H^-} v_\Phi}{2m_{H^\pm}^2} \beta^0_{\gamma\gamma}\left(r_{H^\pm}\right)  + 4\frac{\lambda_{hH^{++}H^{--}} v_\Phi}{2m_{H^{\pm\pm}}^2} \beta^0_{\gamma\gamma}\left(r_{H^{\pm\pm}}\right) \; .
\end{align}
Here $Q_f$ ($I_{3f}$) is the electric charge (third component of the weak isospin) of the SM fermion $f$ running in the loop, $N^c_f$ is the color factor (3 for quarks, and 1 for leptons), $r_X = 4m_X^2/m_h^2$, and the loop functions $\beta^{0,1/2,1}_{\gamma\gamma}$ are collected below: 
%
\begin{align}
&\beta_{\gamma\gamma}^0(x) = -x\left[1-xf(x)\right] \; ,
\\
&\beta_{\gamma\gamma}^{1/2}(x) = 2x\left[1+(1-x)f(x)\right] \; ,
\\   
&\beta_{\gamma\gamma}^1(x) = -\left[2+3x+3x(2-x)f(x)\right]\; ,
\end{align}
with the function $f(x)$ 
defined as
\begin{align}
&f(x) = \begin{cases} 
\left[\sin^{-1}(\sqrt{1/x})\right]^2 & x \ge 1 
\\   
\displaystyle -\frac{1}{4} \left[\log{\frac{1+\sqrt{1-x}}{1-\sqrt{1-x}}} - i\pi\right]^2 & x < 1 
\end{cases} \;.
\end{align}

The trilinear couplings $\lambda_{hH^+H^-}$ and $\lambda_{hH^{++}H^{--}}$ are given by
\begin{align}
\lambda_{hH^+H^-} =& \cos\alpha \Bigg\{\frac{\lambda v_\Phi}{2} \sin^2\beta + \left(\mu - \frac{\lambda_4 v_\Delta}{2\sqrt{2}}  \right) \sin2\beta + \left(\lambda_1+\frac{\lambda_4}{2}\right) v_\Phi \cos^2\beta \Bigg\} \nonumber
\\
& + \sin\alpha \Big\{ \lambda_1 v_\Delta \sin^2\beta + 2(\lambda_2+\lambda_3) v_\Delta \cos^2\beta - \frac{\lambda_4v_\Phi}{2\sqrt{2}}  \sin2\beta \Big\},
\\
\lambda_{hH^{++}H^{--}} = & \lambda_1 v_\Phi \cos\alpha + 2 \lambda_2 v_\Delta \sin\alpha \; ,
\end{align}
where $\beta =\tan^{-1}\left(\sqrt{2}v_\Delta/v_\Phi\right)$ denotes the mixing angle between singly-charged states $\Phi^\pm$ and $\Delta^\pm$.

\subsection{$h\to Z\gamma$}

Similarly, the loop-induced $h\to Z\gamma$ decay rate is given by~\cite{Chen:2013vi, Dev:2013ff}
\begin{align}
\Gamma^{(\rm SM)}_{h \to Z\gamma} = \frac{\alpha_{\rm em} G_F^2 m_W^2 m_h^3}{64\pi^4} \left(1-\frac{m_Z^2}{m_h^2}\right)^3 \left|g^{(\rm SM)}_{hZ\gamma}\right|^2,
\end{align}
where the effective couplings are given by 
\begin{align}
g_{hZ\gamma}^{\rm SM} =& \sum_f \frac{2N^c_f}{\cos\theta_w} Q_f \left(I_{3f} - 2Q_f \sin^2\theta_w \right) \beta^{1/2}_{Z\gamma}\left(r_f, z_f\right)
 + \beta^1_{Z\gamma}\left(r_W, z_W\right) \, ,
\\
g_{hZ\gamma} =& \sum_f \frac{2N^c_f}{\cos\theta_w} Q_f \left(I_{3f} - 2Q_f \sin^2\theta_w \right) g^h_{ff} \beta^{1/2}_{Z\gamma}\left(r_f, z_f\right)
 + g^h_{WW} \beta^1_{Z\gamma}\left(r_W, z_W\right) \nonumber \\
 & \quad 
 - g_{ZH^+H^-} \frac{\lambda_{hH^+H^-} v}{2m_{H^\pm}^2} \beta^0_{Z\gamma} \left(r_{H^\pm}, z_{H^\pm}\right)
 - 2 g_{ZH^{++}H^{--}} \frac{\lambda_{hH^{++}H^{--}} v}{2m_{H^{\pm\pm}}^2} \beta^0_{Z\gamma} \left(r_{H^{\pm\pm}}, z_{H^{\pm\pm}}\right),
\end{align}
with
\begin{align}
g_{ZH^+H^-} = \sin^2 \beta - \frac{2\sin^2\theta_w}{\cos 2\theta_w} \cos^2 \beta \, , \qquad 
g_{ZH^{++}H^{--}} = 2 \, .
\end{align}
The loop functions $\beta^{0,1/2,1}_{Z\gamma}$ are given by 
\begin{align}
&\beta_{Z\gamma}^0(x,y) = \frac{\cos 2\theta_w}{\cos \theta_w} I_1(x,y) \, ,
\\
&\beta_{Z\gamma}^{1/2}(x,y) = I_1(x,y)-I_2(x,y) \, ,
\\
&\beta_{Z\gamma}^1(x,y) = \cos\theta_w \left\{ \left(1+\frac{2}{x}\right) \tan^2\theta_w - \left(5+\frac{2}{x} \right) \right\} I_1(x,y) + 4\cos\theta_w(3-\tan^2\theta_w) I_2(x,y) \, .
\end{align}
with the functions $I_{1,2}(x,y)$ defined as 
\begin{align}
&I_1(x,y) = \frac{xy}{2(x-y)} + \frac{x^2y^2}{2(x-y)^2} \left[ f(x)-f(y) \right] + \frac{x^2y}{(x-y)^2} \left[ j(x)-j(y) \right],
\\
&I_2(x,y) = -\frac{xy}{2(x-y)}\left[ f(x)-f(y) \right],
\end{align}
with 
\begin{align}
&j(x) = \begin{cases} 
\sqrt{x-1} \sin^{-1}(\sqrt{1/x}) & x \ge 1 
\\   
\frac{1}{2} \sqrt{1-x} \left[\log{\frac{1+\sqrt{1-x}}{1-\sqrt{1-x}}} - i\pi\right] & x < 1
\end{cases} \, .
\end{align}

\end{document}